\documentclass[a4paper,10pt,fleqn,usenames,dvipsnames]{article}

\usepackage{lmodern}
\usepackage[T1]{fontenc}
\usepackage{amsmath}
\usepackage{amssymb}

\usepackage{url}
\usepackage{longtable}
\usepackage{multirow}
\usepackage{caption}
\usepackage[labelformat=simple]{subcaption}

\usepackage[nodayofweek]{datetime}
\usepackage{enumerate}
\usepackage{xspace}
\usepackage{xcolor}
\usepackage{graphicx}
\usepackage[pdftex,hidelinks]{hyperref}
\usepackage{bookmark}

\usepackage{float}
\usepackage[capitalise]{cleveref}
\usepackage{makecell}

\newcommand{\SBone}{SB1\xspace}
\newcommand{\SBtwo}{SB2\xspace}
\newcommand{\OBone}{OB1\xspace}
\newcommand{\OBtwo}{OB2\xspace}
\newcommand{\CURTAINs}{\textsc{Curtain}s\xspace}
\newcommand{\CATHODE}{\textsc{Cathode}\xspace}
\newcommand{\CWoLa}{\textsc{CWoLa}\xspace}

\usepackage[backend=biber,giveninits=true,doi=false,sorting=none,style=numeric-comp]{biblatex}
\DeclareFieldFormat[article]{citetitle}{\textit{#1}\isdot}
\DeclareFieldFormat[article]{title}{\textit{#1}\isdot}
\DeclareFieldFormat[unpublished]{citetitle}{\textit{#1}\isdot}
\DeclareFieldFormat[unpublished]{title}{\textit{#1}\isdot}
\DeclareFieldFormat[inproceedings]{citetitle}{\textit{#1}\isdot}
\DeclareFieldFormat[inproceedings]{title}{\textit{#1}\isdot}

\DeclareNameAlias{author}{last-first}

\newif\iffiguresinbody

\usepackage[nonatbib,preprint]{neurips}

\title{CURTAINs for your Sliding Window:\\ Constructing Unobserved Regions by Transforming Adjacent Intervals}
\author{%
  John Andrew Raine, Samuel Klein, Debajyoti Sengupta, Tobias Golling \\
  University of Geneva\\
  \texttt{\{firstname\}.\{surname\}@unige.ch} \\
}

\figuresinbodytrue

\begin{document}
\maketitle

    \begin{abstract}
        We propose a new model independent technique for constructing background data templates for use in searches for new physics processes at the LHC.
        This method, called \CURTAINs, uses invertible neural networks to parametrise the distribution of side band data as a function of the resonant observable. The network learns a transformation to map any data point from its value of the resonant observable to another chosen value.
        Using \CURTAINs, a template for the background data in the signal window is constructed by mapping the data from the side-bands into
        the signal region. 
        We perform anomaly detection using the \CURTAINs background template to enhance the sensitivity to new physics in a bump hunt.
        We demonstrate its performance in a sliding window search across a wide range of mass values.
        Using the LHC Olympics dataset, we demonstrate that
        \CURTAINs matches the performance of other leading approaches which aim to improve the sensitivity of bump hunts, can be trained on a much smaller range of the invariant mass, and is fully data driven.
    \end{abstract}

\section{Introduction}


In the ongoing search for new physics phenomena to explain the fundamental nature of the universe, particle colliders such as the Large Hadron Collider~(LHC) provide an unparalleled window into the energy and intensity frontiers in particle physics.
Searches for new particles not contained within the Standard Model of particle physics~(SM) are a core focus of the physics programme, with the hope to explain observations in the universe which are inconsistent with predictions from the SM, such as dark matter, gravity and the observed matter anti-matter asymmetry.

Many searches at the LHC target specific models built upon theories which contain new particles with particular attributes. However, these searches are only sensitive to a specific model. Due to the vast space of models which could extend the SM, it is unfeasible to perform dedicated searches for all of them. 

One of the cornerstones in the model independent hunt for new physics phenomena at the LHC is the bump hunt, a search for a localised excess on top of a smooth background.
The most sensitive observable for the bump hunt is in an invariant mass spectrum which corresponds to the mass of the particle produced at resonance in particle collisions or decays. 
The invariant mass spectrum comprises non-resonant events, which produce a falling background across all mass values, with particles appearing as bumps on top of this background.
The width of a bump is driven by the decay width of the particle and the detectors resolution.
At the ATLAS and CMS Collaborations \cite{ATLAS:2008xda,CMS:2008xjf} bump hunt techniques are employed to search for new fundamental particles, and were a crucial in the observation of the Higgs boson~\cite{ATLAS:2012yve,CMS:2012qbp}. At the LHCb experiment~\cite{LHCb:2008vvz}, these techniques have also been successfully employed to observe new resonances in composite particles~\cite{LHCb:2020bwg,LHCb:2021vvq,LHCb:2022rpd}.

In a bump hunt, the assumption is made that any resonant signal will be localised. With this assumption, a sliding window fit can be performed using a signal region with a side-band region on either side. As the signal is assumed to be localised, the expected background contribution in the signal region can be extrapolated from the two side-bands. The data in the signal region can be compared to the extrapolated background to test for a significant excess. This test is performed across the whole spectrum by sliding the window. 
In a standard bump hunt, only the resonant observable is used in the sliding window fit to extrapolate the background and test for localised excesses.




However, with the incredible amounts of data collected by the ATLAS and CMS Experiments, and lack of evidence for new particles~\cite{ATLAS:2021ilc,ATLAS:2021zqc,ATLAS:2021yfa,CMS:summary1,CMS:summary2,CMS:summary3}, the prospect of observing a bump on a single spectrum as more data is collected is growing ever more unlikely.
Therefore, attention has turned to using advanced machine learning techniques to improve the sensitivity of searches for new physics, and in particular to improving the reach of the bump hunt approach. Such approaches typically utilise additional discriminatory variables for separating signal from background.

If an accurate background template over discriminatory features can be constructed for the signal region, then the classification without labels method (\CWoLa)~\cite{cwola} can be used to extend the bump hunt.
As shown in Ref.~\cite{cwolabump} the data in the side-bands can be used to construct the template for training the classifier if the discriminatory features are uncorrelated with the resonant variable. 


In this paper we introduce a new method, Constructing Unobserved Regions by Transforming Adjacent Intervals (\CURTAINs).
By combining invertible neural networks (INNs) with an optimal transport loss~\cite{villani2009optimal,rubner2000earth,cuturi2013sinkhorn}, we learn the optimal transport function between the two side-bands, and use this trained network (henceforth, referred to as the 'transformer') to construct a background template  by transforming the data from each side-band into the signal region. 

\CURTAINs is able to construct a background template for any set of observables, thus classifiers can be constructed using strongly correlated observables.
These variables provide additional information and are often the best variables for discriminating signal from background and therefore increase the sensitivity of the search.
Furthermore, \CURTAINs is a fully data driven approach, requiring no simulated data.


In this paper we apply \CURTAINs to a search for new physics processes in dijet events produced at the LHC and recorded by a general purpose detector, similar to the ATLAS or CMS experiments. We demonstrate the performance of this method using the R\&D dataset provided from the LHC Olympics (LHCO) \cite{lhco4536377}, a community challenge for applying anomaly detection and other machine learning approaches to the search for new physics~\cite{Kasieczka:2021xcg}.

We demonstrate that \CURTAINs can accurately learn the conditional transformation of background data given the original and target invariant mass of the events. Classifiers trained using the background template provided by \CURTAINs outperform leading approaches, and the improved sensitivity to signal processes matches or improves upon the performance in an idealised anomaly detection scenario.


Finally, to demonstrate its applicability to a bump hunt and observing potential new signals, we apply the \CURTAINs method in a sliding window approach for various levels of injected signal data and show that excesses above the expected background can be observed without biases or spurious excesses in the absence of a signal process.




\section{The Dataset}
The LHCO R\&D dataset comprises two sets of labelled data. Background data from the Standard Model is produced through QCD dijet production, and signal events from the decay of a new particle to two lighter new particles, which each decay to two quarks $W^\prime\rightarrow X\left(\rightarrow q\bar{q}\right)Y\left(\rightarrow q\bar{q}\right)$, where the three new particles have mass $m_{W^\prime} = 3.5$~TeV, $m_{X} = 500$~GeV, and $m_{Y} = 100$~GeV.
Both samples are generated with \texttt{Pythia~8.219}~\cite{Sjostrand:2007gs} and interfaced to \texttt{Delphes 3.4.1}~\cite{deFavereau:2013fsa} for the detector simulation. The reconstructed particles are clustered into jets using the anti-$k_{t}$ algorithm~\cite{Cacciari:2008gp} using the \texttt{FastJet} package~\cite{Cacciari:2011ma}, with a radius parameter $R=1.0$. Each event is required to have two jets, with at least one jet passing a cut on its transverse momentum $p_{\mathrm{T}}^J > 1.2$~TeV to simulate a jet trigger in the detector.

In total 1~million QCD dijet events and 100 thousand signal events are generated. \CURTAINs uses all the QCD dijet events as the standard background sample, and in addition doped samples are constructed using all the QCD events and a small number of events from the signal sample from the 100,000 available $W^{\prime}$ events.
The standard benchmark datasets used to asses the performance of \CURTAINs comprise the full background dataset, with 0, 500, 667, 1000 or 8000 injected signal events.

All event observables are constructed from the two highest $p_{\mathrm{T}}$ jets, with the two jets ordered by their invariant mass, such that $J_{1}$ has $m_{J_1} > m_{J_2}$.
The studied features include the base set of variables introduced in Ref.~\cite{Nachman:2020lpy} and applied in Ref.~\cite{cathode},
\begin{equation*}
     m_{JJ},\ m_{J_1},\ \Delta m_{J} = m_{J_1} -  m_{J_2},\ \tau_{21}^{J_1},\ \tau_{21}^{J_2},
\end{equation*}
where $\tau_{21}$ is the $n$-subjettiness ratio of a large radius jet~\cite{nsubjettiness}, measuring whether a jet has underlying substructure more like a two prong or one decay, and $m_{JJ}$ is the invariant mass of the dijet system.
As an additional feature we include
\begin{equation*}
    \Delta R_{JJ},
\end{equation*}
which is the angular separation between the two jets in $\eta-\phi$ space. 
This additional feature is included as it can bring additional sensitivity to some signal models.
Furthermore, it is strongly correlated with the the resonant feature, $m_{JJ}$, and so including it when training the transformer and classifier provides a stringent test of the \CURTAINs method.

The width of the signal region in the sliding window is set to 200~GeV by default, with 200~GeV wide side-bands either side. In this paper, we simplify the sliding window approach by shifting the window by 200~GeV such that there is no overlap between signal regions. This would reduce the sensitivity for cases where the signal peak falls on the boundary of the signal region. We avoid this by defining our bins such that the signal is centred within a signal region. Where the signal is unknown overlapping windows would need to be employed, with a strategy in place to avoid selecting the same data twice in the final analysis.
The turn on in the dijet invariance mass spectrum caused by the trigger requirements of both jets is removed by only performing the sliding window scan with signal regions above 3.0~TeV. The full range used for the sliding window scan is up to a dijet invariant mass of 4.6~TeV.

To evaluate the performance of classifiers using this dataset, a $k$-fold procedure with five folds is employed, using three fifths of the dataset for training, one fifth for validation and one fifth as a hold out set per fold. No optimisation is performed on the hold out sets, and all optimisation criteria are satisfied using the validation set per fold. This ensures all available data are used in a statistical analysis,
which is even more crucial in data driven approaches, where statistical precision is key in the search for new physics.
The remaining 92,000 signal events not used to construct the doped datasets are used to evaluate the classifier performance, maximising the statistical precision.

\section{Method}
\subsection{\CURTAINs}

In \CURTAINs conditional invertible neural networks~(cINNs)~\cite{ardizzone2019analyzing,ardizzone2019guided} are employed to transform data points from an input distribution to those from the target distribution. The transformation is conditioned on a function $f$ of the resonant feature $m_{JJ}$ of the input and target data points.
Unlike flows~\cite{rezende2016variational,2021flows}, which use the exact maximum likelihood of transforming data to a desired distribution, usually a multivariate normal distribution, we use an optimal transport loss to train the network to transform data between the two desired distributions. 

As the cINN can be used in both directions, the inputs to the conditional function are referred to as the lower and higher values $m_{JJ}^{low}$ and $m_{JJ}^{hi}$. In the case of a forward pass through the network, $m_{JJ}^{low}$ are the true values of the input data, with $m_{JJ}^{hi}$ the target values, and vice versa in the case of an inverse pass.
Furthermore, instead of training the cINN in only the forward direction, we iterate between both the forward and inverse directions to ensure better closure between the output and target distributions and to prevent a bias towards transformations in one direction. 

Several different network architectures for the transformer were studied in the development of \CURTAINs. The transformers presented in this paper are built on the invertible transformations introduced in Ref.~\cite{durkan2019neural} which use rational-quadratic~(RQ) splines, which are found to be very expressive and easy to train. The conditioning function $f$ is chosen to be
\begin{equation}
    f\left(m_{JJ}^{low}, m_{JJ}^{hi}\right) = m_{JJ}^{hi} - m_{JJ}^{low}.
\end{equation}
The features which are to be transformed determine the input and target dimensions of the \CURTAINs transformer.

To train the network, batches of data are drawn from the low-mass side-band \SBone and the high-mass side-band \SBtwo. The data from \SBone is first fed through the network in a forward pass, conditioned using $m_{JJ}^{low}$ and $m_{JJ}^{hi}$, with target values for each event assigned by randomly pairing the masses drawn from each side-band.
The loss between the transformed data and target data is calculated using the sinkhorn divergence \cite{cuturi2013sinkhorn} across the whole batch, in order to measure the distance between the distributions of the two sets of data. The gradient of this loss is used to update the network weights.
In the next batch the data from \SBtwo is fed through the network in an inverse pass and the same procedure is performed.
This alternating procedure is repeated throughout the training.
A schematic overview of the \CURTAINs transformer model is shown in \cref{fig:curtains_model}.

\iffiguresinbody
\begin{figure}[htbp]
    \centering
    \includegraphics[width=0.75\textwidth]{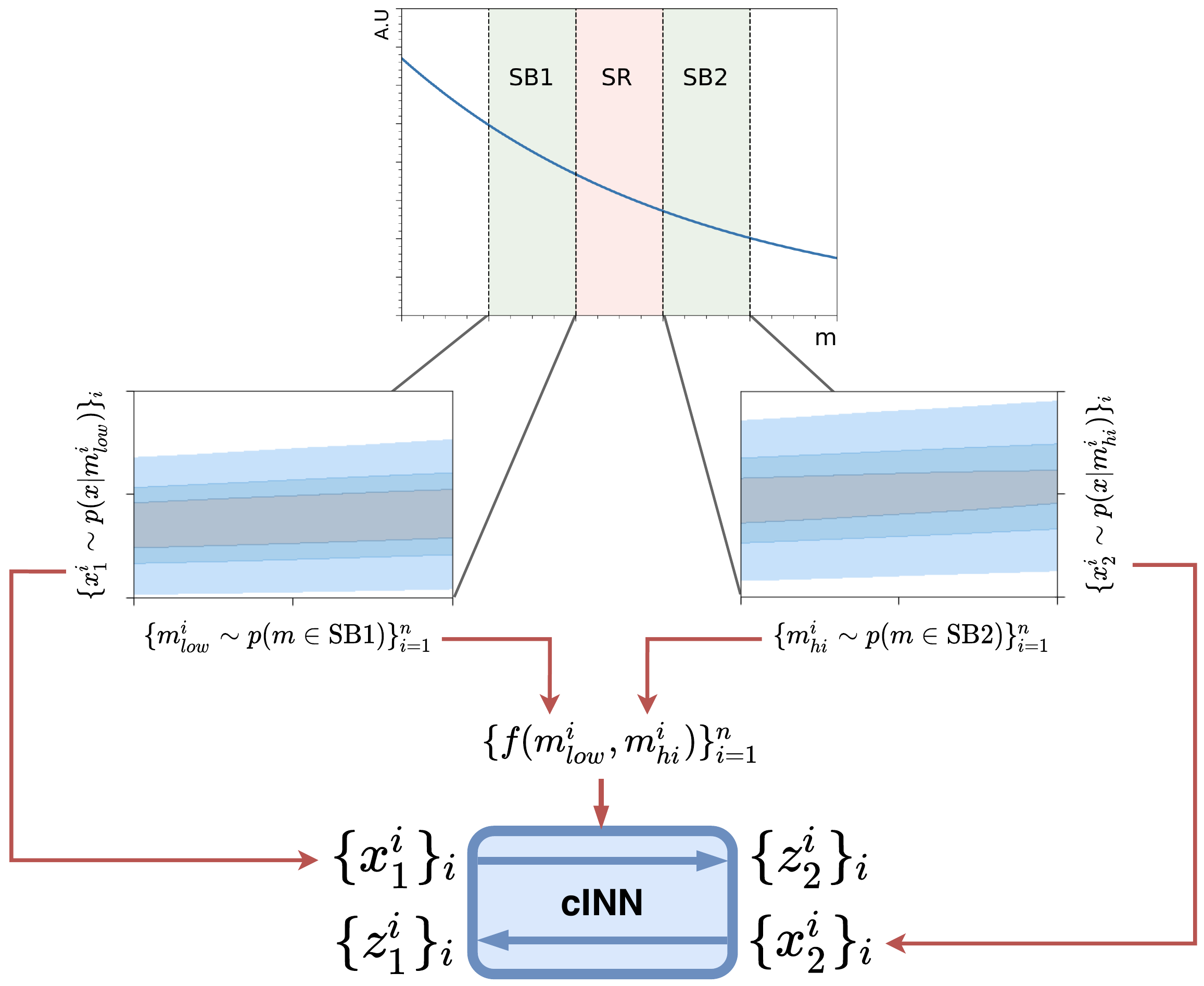}
    \caption{A schematic overview of the \CURTAINs model. 
    A feature $x$ is correlated with $m$, as can be seen from the 2D contour plots for each side-band in blue.
    Samples $\{x_i^1\}_{i=1}^n$ and $\{x_i^2\}_{i=1}^n$ of batch size $n$ are drawn randomly from the two side-bands. 
    In the forward pass the samples from \SBone, $\{{x_1}\}_{i=1}^n$, are passed through the conditional INN where each sample $x_1^i$ is conditioned on $f\left(m^i_{low},m^i_{hi}\right)$, producing the set ${\{z_2^i}\}_{i=1}^n$. The cost function is defined as the distance between this output and the sample from \SBtwo $\{{x_2^i}\}_{i=1}^n$.
    In the inverse pass the roles of each side-band are exchanged. In applying the model, any value for $m$ can be chosen as long as the correct inverse or forward pass is applied.
    }
    \label{fig:curtains_model}
\end{figure}
\fi

With this training procedure the optimal transport function is not exactly derived as the conditional information is only implicit and a transformed event will not necessarily be paired to the event with the mass to which it was mapped in the loss calculation. However, after training the network we observe that the learned transformation is a good approximation of the true optimal transformation.

In order to improve the closure of the transformed data to regions other than the the side-bands, an additional training step is performed. After an epoch of training the network between \SBone and \SBtwo, each side-band itself is split into two equal width sub side-bands. The network is then trained for an epoch of each intra side-band training, following the same procedure as for the inter side-band training.
Although not necessary for the \CURTAINs method, this extra step is performed in order to extend the range of values of the conditional information used to train the network. Instead of having a minimum value of $f\left(m_{JJ}^{low},m_{JJ}^{hi}\right)$ equal to the width of the signal region separating \SBone and \SBtwo, its minimum values is now zero. This ensures that the conditioning variables used to map data to the signal region always lie in the distribution of values used during training.

The \CURTAINs transformer is trained for 1000 epochs with a batch size of 256 using the \texttt{Adam} optimiser \cite{kingma2017adam}. A cosine annealing learning rate schedule is used with an initial learning rate of $10^{-4}$.
A typical training time of 6~hours using an NVIDIA$^\textrm{\textregistered}$ 3080 RTX GPU is required for a central window encompassing approximately $10^5$ samples across the two side-bands.

The \CURTAINs transformers are trained separately for each step in the sliding window, using all the available data in the side-bands.
In order to construct a background template in another region, all the data from \SBone and \SBtwo are transformed in either a forward or inverse pass to mass values sampled from the target window.
To create the background template in the signal region, the data from \SBone (\SBtwo) are transformed to values of $m_{JJ}$ corresponding to the signal region in a forward (inverse) pass with the \CURTAINs transformer. These two transformed datasets are combined to create the background template in the signal region.

In the case of validating the \CURTAINs transformer, the side-band data can be transformed to a target window with the same width as the signal region but going in the opposite direction in $m_{JJ}$, defining outer-band regions for \SBone (\OBone) and \SBtwo (\OBtwo).
These regions can be used to validate and tune the \CURTAINs method in a real world setting.
The five bands of one sliding window are illustrated in \cref{fig:windows}, with the depicted signal region centred on the invariant mass of the injected signal.
In the studies presented in this paper the width of the side-bands and validation regions is set to 200~GeV by default, unless otherwise specified.
To increase the statistics of the constructed datasets the transformer can be applied many times to the same data with different $m_{JJ}$ target values in each pass.

The hyperparameters and architecture of the \CURTAINs transformer were optimised in a grid search by measuring the agreement between data transformed into the two outer-band regions from the two side-bands for one step of the sliding window without any doping of signal events. 
The agreement is measured by training a classifier to separate the two datasets and ensuring the Receiver Operator Characteristic (ROC) curve had a linear response with an area under the curve close to 0.5, which suggests the network was unable to differentiate between real and transformed data in this region. The optimal \CURTAINs transfomer is made up of eight stacked RQ spline coupling layers. Each coupling layer is constructed from three residual blocks each of two hidden layers of 32 nodes with $\textsc{Leaky ReLU}$ activations, resulting in an output spline with four bins.
The \texttt{n-flows} package~\cite{conor_durkan_2020_4296287} is used to implement the network architecture in Pytorch~1.8.0~\cite{NEURIPS2019_9015}.
These settings are then used to train all \CURTAINs transformers for each step of the sliding window, and for all doping levels.

\iffiguresinbody
\begin{figure}[h]
    \centering
    \includegraphics[width=0.55\textwidth]{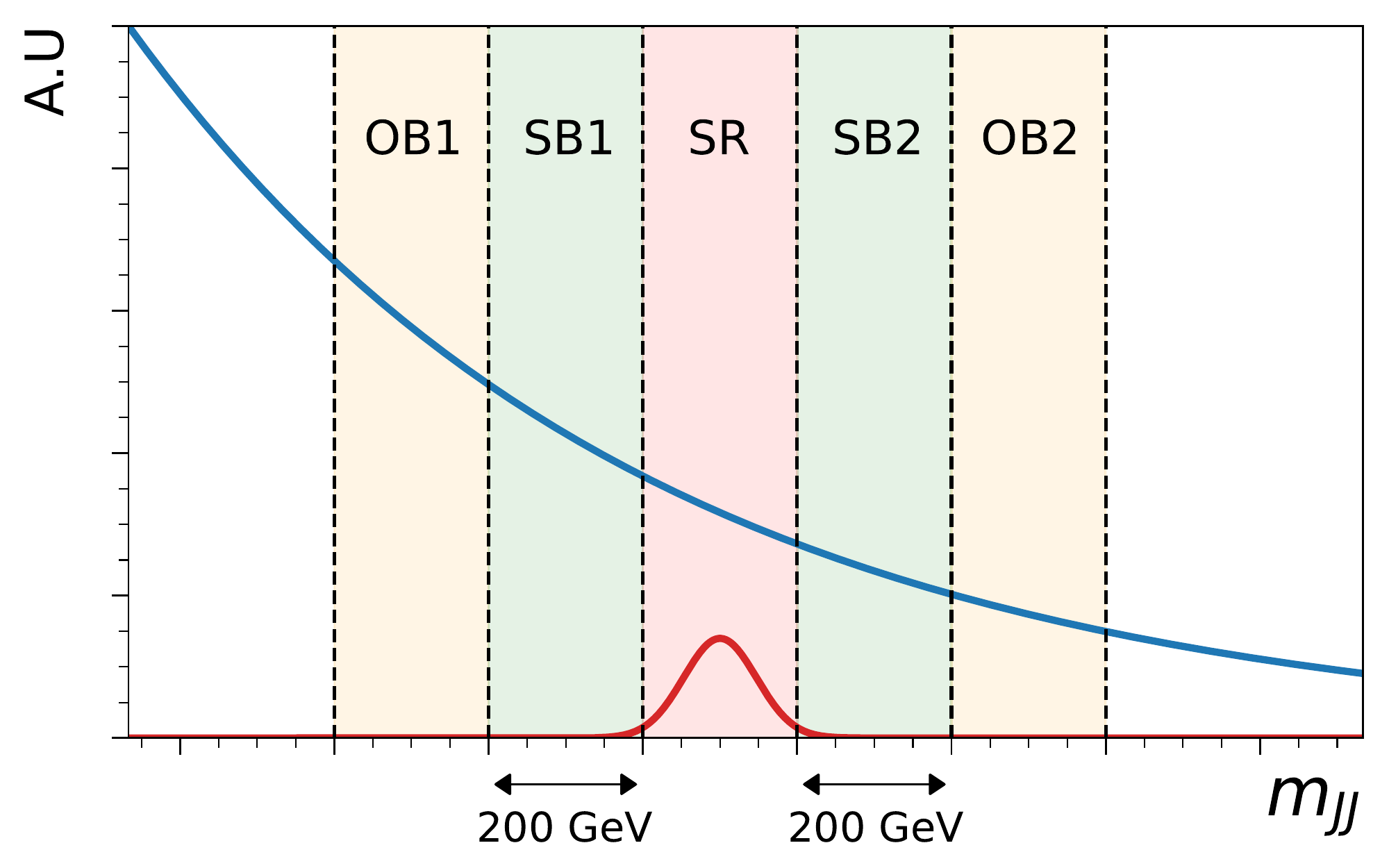}
    \caption{Schematic showing the relative locations of the two side-bands (\SBone and \SBtwo), the signal region (SR) and the two outer-bands (\OBone and \OBtwo) on the resonant observable $m_{JJ}$. In this example, the non-resonant background is shown as a falling blue line, and the signal region is centred at 3.5~TeV, corresponding to the mass of the injected signal, shown not to scale in red.}
    \label{fig:windows}
\end{figure}
\fi

\subsection{Mass Fitting and Sampler}

In order to sample target values for the \CURTAINs transformer and not be biased to the presence of any excess of events in the signal region, the distribution of the resonant feature in the signal region needs to be extrapolated from the side-band data.
Here we model the QCD dijet background with the functional form
\begin{equation}
    f(z)=p_1 (1-z)^{p_2} z^{p_3},
\end{equation}
where $z = m_{JJ}/\sqrt{s}$ with the centre of mass energy of the collision $\sqrt{s}=$13~TeV. The parameters $p_1$, $p_2$, and $p_3$ are obtained from an unbinned fit to the side-band data in using the \texttt{zfit} package~\cite{ESCHLE2020100508}. This Ansatz has been used previously in analyses performed at the LHC~\cite{ATLAS:2015nsi} and is similar to that used in more recent searches with the omission of the last free parameter~\cite{ATLAS:2019fgd,CMS:2018mgb}.
Once fit to the side-band data, the learned parameters are used in the PDF from which to sample target $m_{JJ}$ values for the transformer.

\subsection{Anomaly Detection}

Once the background data has been transformed into the signal region from the side-bands, it is possible to use them as the background template to test for the presence of signal in the data from the region.
There are several approaches which could be used for anomaly detection with the data transformed with the \CURTAINs method, however in this paper we will focus on the \CWoLa classifier, as applied also in Refs.~\cite{cathode,cwolabump,Benkendorfer:2020gek} on this dataset.

For a \CWoLa classifier, it can be shown that the performance of a classifier trained on two sets of data, each containing a different mixture of signal and background data will result in the optimal classifier trained on pure sets of signal and background data. Here, we assume our transformed data represents a sample of pure background events, and test the hypothesis that in our signal region data there is a mixture of signal and background data.
In the presence of signal events in the signal region, the classifier will be able to separate the signal region data from the background template, with the true signal events having higher classification scores than the true background data. By applying a cut on the classifier output to reject a given fraction of the background, calculated from the scores of the background template, the significance of the signal events can be enhanced.

In cases where there is signal contamination in at least one of the side-bands of the sliding window, the background template constructed with \CURTAINs will also contain a non-zero signal to background fraction. With the assumption that the signal is localised, and the bin widths are not too small, the relative fraction of signal in the signal region will be different from the background template. As such, the \CWoLa method will still be able to approach the performance of the ideal classifier. The background template provided by \CURTAINs will have a lower signal to background ratio than the signal region in at least one step of the sliding window, and in this bin an excess can be expected. 

In the event of the signal being fully localised within a side-band, this will result in the opposite labels being used in the training of the \CWoLa classifier with regards to which dataset contains the higher fraction of signal. After applying a cut on the classifier a slight reduction in events with respect to the prediction could therefore be expected. However, in practise we observe no significant deviation with the dataset under consideration.

The values used as acceptance thresholds on the output of the classifier are independently determined for each classifier in the signal regions across all sliding windows and levels of doping.
These cuts are used to enhance the sensitivity to the presence of signal data in each window of the fit.
The amount of data which remains after the cut can be compared to the expected background, determined by taking the total number of data in each signal region multiplied by the background rejection factor.
In the presence of a signal, a significant excess of data will be observed above the expected background.


A further test of the performance when using the \CURTAINs method is to compare against three benchmark classifiers. The first is a fully supervised classifier, trained with knowledge of which events were from the signal process and which were QCD background.
Two further classifiers, the idealised classifiers, are trained in the same manner as with the \CURTAINs background template, except that the background template comprises true background data from the signal region itself.

Both the supervised and idealised classifiers are only trained for the window in which the signal region is aligned with the peak of the signal data. The supervised classifier provides an upper bound on the achievable performance on the dataset. The idealised classifier sets the target level of performance which can be achieved with a perfect background template, and can be used to validate the performance of \CURTAINs for use in anomaly detection.

All the classifiers for all signal regions and all levels of signal doping share the same architecture and hyperparameters.
The classifiers used in this paper have been chosen as they are robust to changes in datasets and initial conditions, in particular when using $k$-fold training and low training statistics. The classifiers are constructed from multilayer perceptrons with three hidden layers with 32 nodes and $\textsc{ReLU}$ activations. The classifiers are trained for 20 epochs using the \texttt{Adam} optimiser with a batch size of 128, and an initial learning rate of $0.001$ which anneals to zero following a cosine curve over 20 epochs.

\section{Comparison to other work}
Our method is one of several approaches with aims to enhance the sensitivity to new physics processes coming from the resonant production of a new particle using machine learning \cite{cwolabump,anode,cathode,Andreassen:2020nkr,Benkendorfer:2020gek}.

In comparison to the \CATHODE method introduced in Ref.~\cite{cathode}, which is one of the current best anomaly detection methods for resonant signals using the \CWoLa approach, our method shares some similarities but differs on key points.
Although both approaches make use of INNs, \CURTAINs does not train a flow with maximum likelihood but instead uses an optimal transport loss in order to minimise the distance between the output of the model and the target data, with the aim to approximate the optimal transport function between two points in feature space when moving along the resonant spectrum.
As a result, \CURTAINs does not generate new samples to construct the background template, but instead transforms the data in the side-bands to equivalent datapoints with a mass in the signal region. This approach avoids the need to match data encodings to an intermediate prior distribution, normally a multidimensional gaussian distribution, which can lead to mismodelling of underlying correlations between the observables in the data if the trained posterior is not in perfect agreement with the prior distribution. 
The \CATHODE method has no regularisation on the model's dependence on the resonant variable, and this dependence is non trivial, so extrapolating to unseen datapoints -- such as the signal region -- can be unreliable. 
In contrast, the \CURTAINs method can be constructed such that at evaluation the conditioning variable is never outside of the values seen from the training data.

Furthermore, in comparison to \CATHODE, \CURTAINs is designed to be trained only in the sliding window with all information extracted over a narrow range of the resonant observable, as is standard in a bump hunt. This means \CURTAINs is less sensitive to effects from multiple resonances on the same spectrum, and is not dominated by areas of the distribution with more data. Furthermore, thanks to the optimal transformation learned between the side-bands, it can also be applied to transform side-band data into additional validation regions and not just to construct the background template in the signal region.

In contrast to the methods proposed in Ref.~\cite{Andreassen:2020nkr} (\textsc{Salad}) and Ref.~\cite{Benkendorfer:2020gek} (\textsc{SA-CWoLa}), \CURTAINs does not rely on any simulation and is a completely data-driven technique. In \CURTAINs the side-band data is able to be transformed directly into the signal region, instead of deriving a reweighting between the data and simulated data from the side-bands, which is subsequently applied to transform the simulated data in the signal region into a background template.
Due to the resampling of the value of the resonant observable, \CURTAINs is also able to produce a background template with additional statistics, rather than being limited by the number of events in the signal region from the simulated sample.

There are also a wide range of approaches looking for new physics that do not rely on resonant signals. Many techniques are built on autoencoders~\cite{Aguilar-Saavedra:2017rzt,Hajer:2018kqm,Heimel:2018mkt,Farina:2018fyg,Cerri_2019,Roy:2019jae,Blance_2019,Jawahar:2021vyu}, looking to identify uncommon events or objects. These models are subsequently used to reject SM-like processes in favour of potential new physics.
Other approaches are motivated from the ratio of probability densities and directly measure a test statistic from the comparison of a sample of events with respect to a set of reference distributed events \cite{D_Agnolo_2019,De_Simone_2019,Letizia:2022xbe,DAgnolo:2019vbw}.
A comparison of a wide range of methods is also performed in Ref.~\cite{Kasieczka:2021xcg}, which summarises a community challenge for anomaly detection in high energy physics.

\section{Results}
\subsection{Validating \CURTAINs transformer}
The first test of performance in \CURTAINs is to demonstrate that the transformation learned between the two side-bands is accurate, and further to determine whether the learned transformation can extrapolate well to the validation regions.
As Monte Carlo simulation is being used for the studies, we can control the composition of the samples in the studies. The performance of the approach is evaluated using a sample containing only background data, as well as various levels of signal doping. The same model configuration is used for all samples, and the sliding window is chosen such that it is centred on the true signal peak with a signal region width of 200~GeV.

The input features and their correlations for the input, target and transformed data distributions are shown in \cref{fig:curtains_nosignal_SBs} for the two side-bands trained in the case of no signal, and in \cref{fig:curtains_nosignal_OBs} for the two validation regions. 
As can be seen, the transformed data distributions are well reproduced with the \CURTAINs approach. The ability of \CURTAINs to handle features which are strongly correlated with $m_{JJ}$ can be seen from the agreement of the $\Delta R_{JJ}$ distributions between \SBone and \SBtwo, which exhibit very different shapes. 

\iffiguresinbody
\begin{figure}[htpb]
    \centering
    \includegraphics[width=\textwidth]{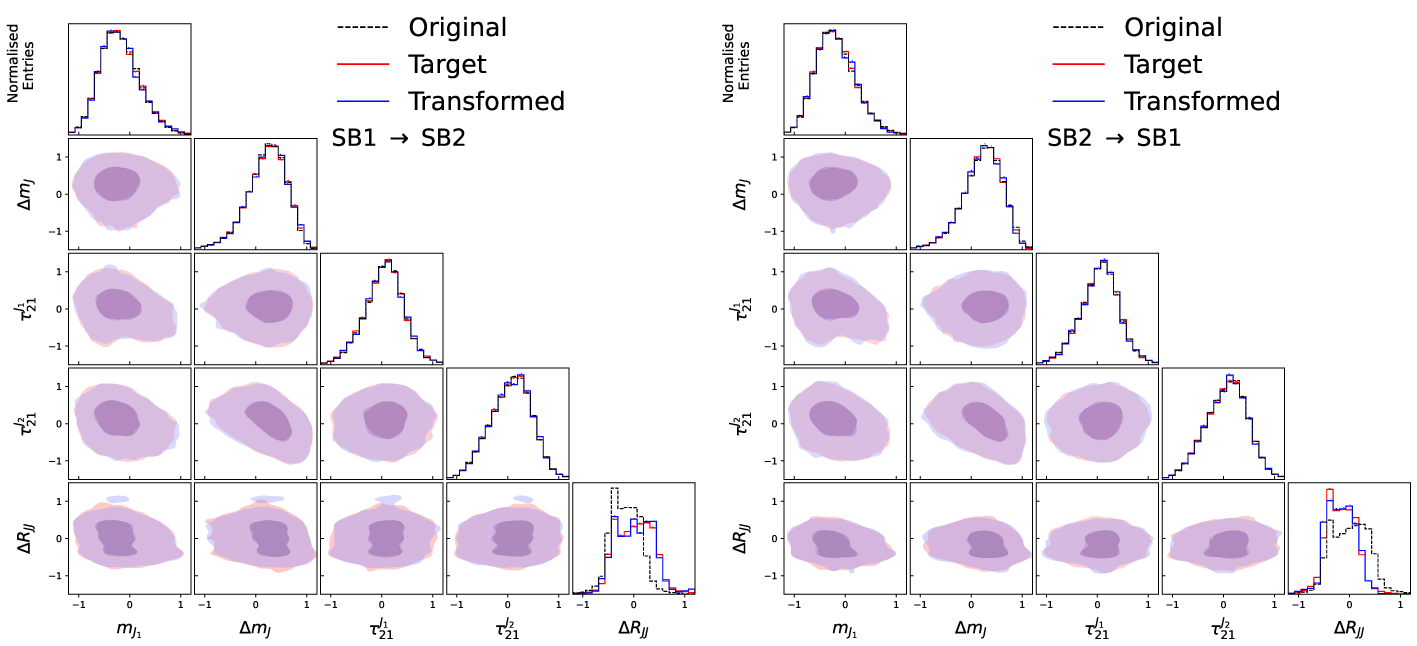}
    \caption{Input, target and transformed data distributions for the base variable set with the addition of $\Delta R_{JJ}$, for transforming data from \SBone to \SBtwo (left) and \SBtwo to \SBone (right), with the model trained on \SBone (3200~$\leq m_{JJ} <$~3400~GeV) and \SBtwo (3600$~\leq m_{JJ} <$~3800~GeV). The data from \SBone (\SBtwo) is transformed with a forward (inverse) pass of the \CURTAINs model into the target region.
    The diagonal elements show the individual features with the off diagonal elements showing a contour plot between the two observables for the transformed and trained data.}
    \label{fig:curtains_nosignal_SBs}
\end{figure}
\fi

\iffiguresinbody
\begin{figure}[htpb]
    \centering
    \includegraphics[width=\textwidth]{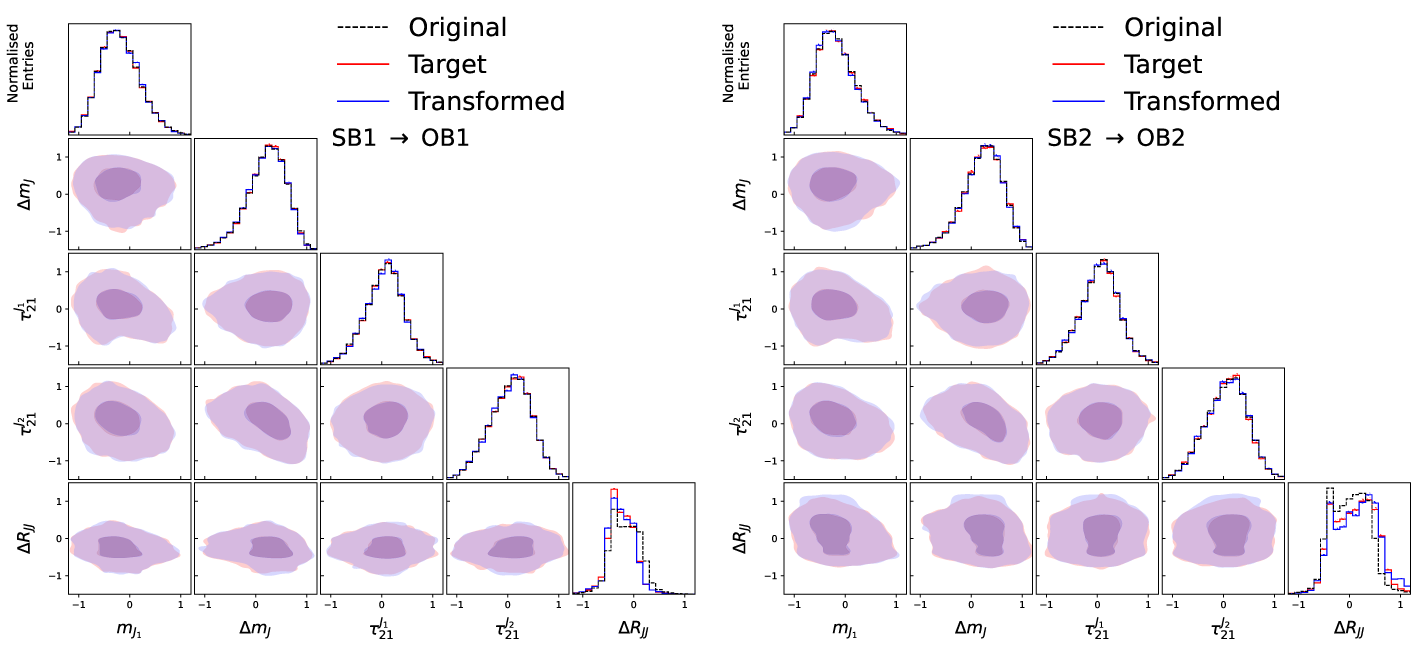}
    \caption{Input, target and transformed data distributions for the base variable set with the addition of $\Delta R_{JJ}$, for transforming data from \SBone to \OBone (left) and \SBtwo to \OBtwo (right), with the model trained on \SBone (3200~$\leq m_{JJ} <$~3400~GeV) and \SBtwo (3600$~\leq m_{JJ} <$~3800~GeV), with \OBone and \OBtwo defined as 200~GeV wide windows directly next to \SBone and \SBtwo away from the signal region.
    The data from \SBone (\SBtwo) is transformed with an inverse (forward) pass of the \CURTAINs model into the target region.
    The diagonal elements show the individual features with the off diagonal elements showing a contour plot between the two observables for the transformed and trained data.}
    \label{fig:curtains_nosignal_OBs}
\end{figure}
\fi

In the case of no signal being present, we can also verify whether the background template constructed by transforming data from the side-bands with \CURTAINs matches the target data in the signal region. The performance of the \CURTAINs method can be seen in \cref{fig:curtains_nosignal_SR}, with the transformed data closely matching the data distributions and correlations.

\iffiguresinbody
\begin{figure}[htpb]
    \centering
    \includegraphics[width=0.66\textwidth]{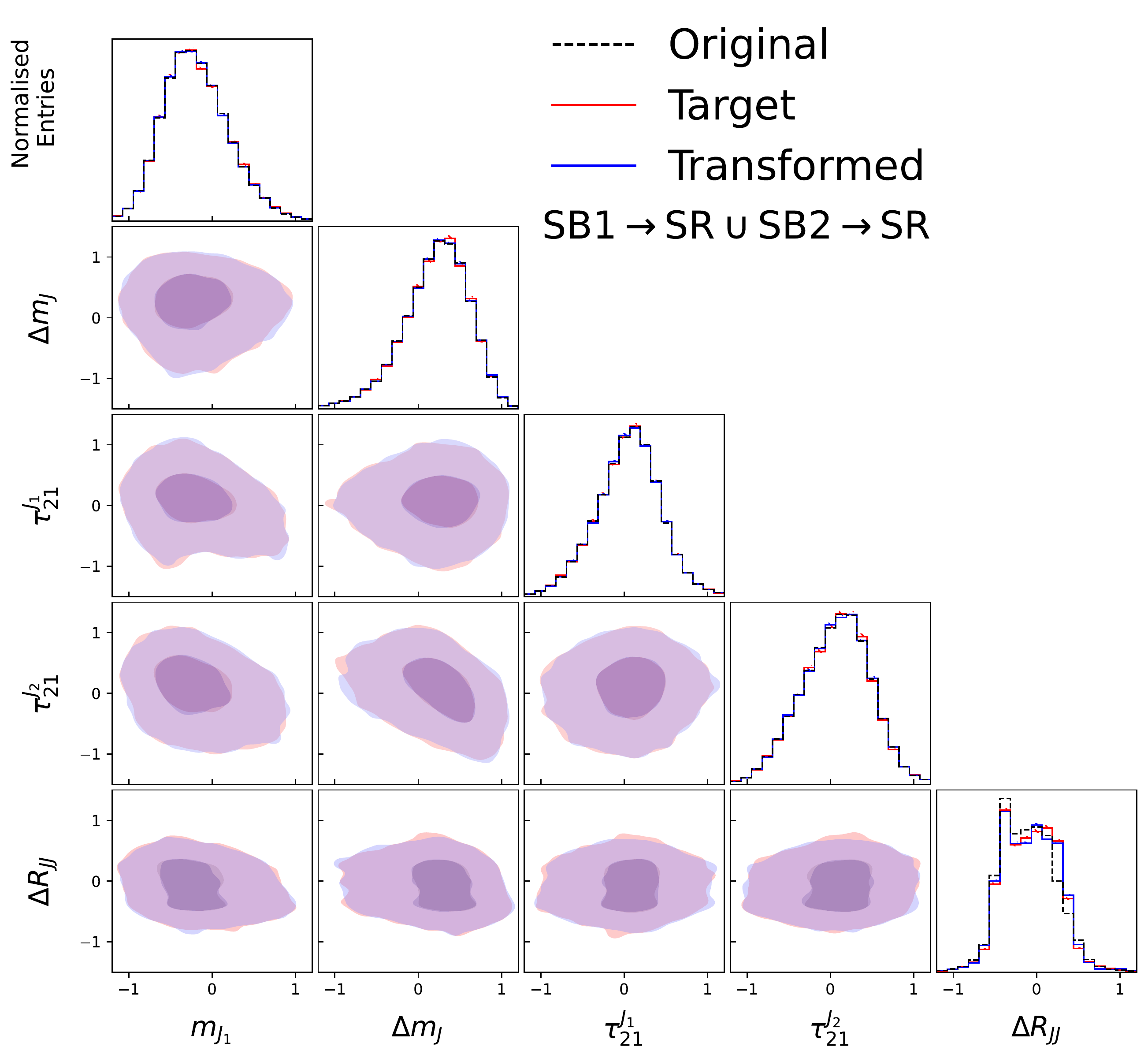}
    \caption{Input, target and transformed data distributions for the base variable set with the addition of $\Delta R_{JJ}$, for transforming data from \SBone and \SBtwo to the signal region to create the background template, with the model trained on \SBone (3200~$\leq m_{JJ} <$~3400~GeV) and \SBtwo (3600$~\leq m_{JJ} <$~3800~GeV). The data from \SBone (\SBtwo) is transformed with a forward (inverse) pass of the \CURTAINs model into the target region.
    The diagonal elements show the individual features with the off diagonal elements showing a contour plot between the two observables for the transformed and trained data.}
    \label{fig:curtains_nosignal_SR}
\end{figure}
\fi

To quantify the level of agreement between the transformed distributions and the target data, classifiers are trained to separate the two datasets, and the area under the ROC curve is measured. The level of agreement between the \CURTAINs transformed data and target data can be seen for several levels of signal doping in \cref{tab:curtains_aucs}. We can see that \CURTAINs has very good agreement with the target distribution in all signal regions and in all cases is seen to be better than in the validation region. The reduced performance in \OBone and \OBtwo is a result of the transformer extrapolating outside of the trained sliding window. This demonstrates their ability to be used for validating the \CURTAINs method and future classification architectures.

\iffiguresinbody
\begin{table}[htpb]
    \caption {Quantitative agreement between the data distributions of the transformed data and the target data as measured by the AUC of the ROC curve trained on the two samples, as measured for various levels of signal doping with a 200 GeV wide signal region.}
    \label{tab:curtains_aucs}
    \centering
    \begin{tabular}{r c c c c | c}
        \hline
        \hline
        & \SBone\!\!$\rightarrow$\SBtwo & \SBtwo\!$\rightarrow$\SBone & \SBone\!\!$\rightarrow$\OBone & \SBtwo\!$\rightarrow$\OBtwo &  \thead{\SBone\!\!$\rightarrow$SR \\ $\cup$\\ \SBtwo\!$\rightarrow$ SR}\\
        \hline
        
        0 signal    & 0.504 & 0.504 & 0.519 & 0.512 & 0.509   \\
        500 signal  & 0.503 & 0.503 & 0.518 & 0.506 & 0.506   \\
        667 signal  & 0.505 & 0.504 & 0.516 & 0.514 & 0.505   \\
        1000 signal & 0.499 & 0.502 & 0.520 & 0.502 & 0.512   \\
        8000 signal & 0.508 & 0.511 & 0.523 & 0.521 & 0.522   \\
        \hline\hline
    \end{tabular}
\end{table}
\fi

\subsection{Application to anomaly detection}
To demonstrate the performance of \CURTAINs to produce a robust background template, the sliding window is centred on the resonant mass of the signal events, and the performance of the \CWoLa classifier is compared against a background template produced using the \CATHODE method. 
The signal region width is set to 400~GeV to contain the majority of the signal events, resulting in 120,000 background events.
The background template is produced with oversampling, with a total of nine times the number of expected events in the signal region.
Two comparisons to \CATHODE can be performed, one using the same training windows as for \CURTAINs, which we refer to as \CATHODE~(local), and one using the full invariant mass distribution outside the signal region, as presented in Ref.~\cite{cathode}, which we refer to as \CATHODE~(full).

For reference, the methods are compared to a classifier trained using an idealised background template and to a fully supervised classifier.
The idealised background template constructed using true background events from the signal region, and the supervised classifier is trained to separate the signal data from the background data using class labels.
To construct the idealised background dataset we either use an equal number of background data as there are in the signal region (Eq-Idealised) to measure the performance assuming we had access to a perfect model of the background data, or the same number of data points as are produced with the \CURTAINs and \CATHODE approaches (Over-Idealised), which should approach the best possible performance for models which can oversample.

The performance of the classifiers with the different methods are shown for the doped sample with 3000 injected signal events (of which 2214 are in the signal region) in \cref{fig:ROCandSIC}, comparing the background rejection as a function of signal efficiency and the significance improvement
as a function of the background rejection. In order to maintain a fair comparison to the Eq-Idealised classifier, which requires true background data from the signal region for the background template, only half of the available data in the signal region is used for training with the $k$-fold strategy for all other approaches.
The maximum significance improvement is shown for a wide range of doping levels in \cref{fig:dopinglevelscan}. This metric is a good measure of performance for anomaly detection, rather than the area under the ROC curve, as it translates to the expected performance gain when applying an optimal cut on a classifier.

\iffiguresinbody
\begin{figure}[htpb]
    \centering 
    \includegraphics[width=\textwidth]{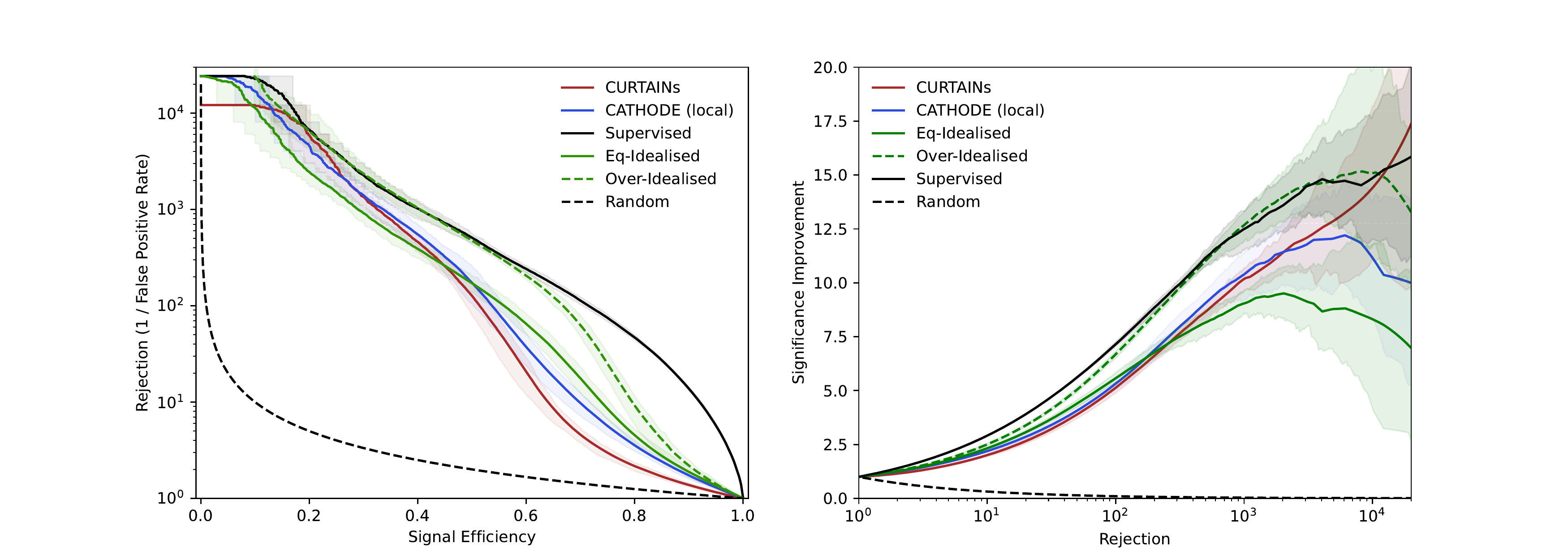}
    \caption{Background rejection as a function of signal efficiency (left) and signal improvement as a function of background rejection (right) for the different background template models ($\CURTAINs$ - red, $\CATHODE$ - blue, Eq-Idealised - green, Over-Idealised - dashed green) and a fully supervised classifier (black).
    The sample with 3000 injected signal events in used to train all classifiers in the signal region 3300~$\leq m_{JJ} <$~3700~GeV.
    The solid lines show the mean value of fifty classifier trainings with different random seeds. The uncertainty encompasses 68\% of the runs either side of the mean. 
    }
    \label{fig:ROCandSIC}
\end{figure} 
\fi

\iffiguresinbody
\begin{figure}[htpb]
    \centering
    \includegraphics[width=0.6\textwidth]{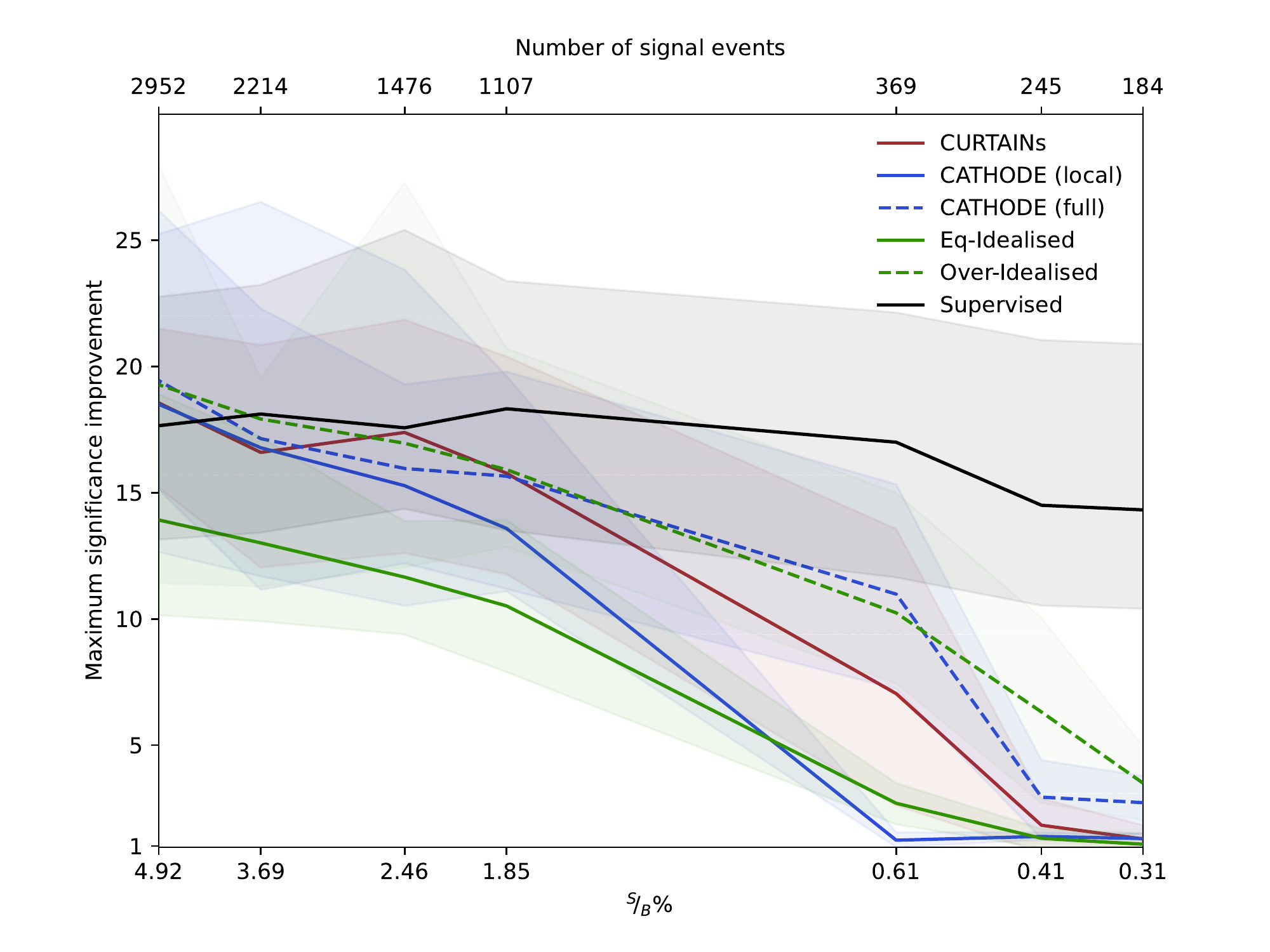}
    \caption{The significance improvement as a function of decreasing signal purity (raw signal events)  for the different background template models ($\CURTAINs$ - red, $\CATHODE$~(local) - blue, Eq-Idealised - green, Over-Idealised - dashed green) and a fully supervised classifier (black). All classifiers trained in the signal region 3300~$\leq m_{JJ} <$~3700~GeV for varying levels of signal doping.
    The solid lines show the mean value of fifty classifier trainings with different random seeds. The uncertainty encompasses 68\% of the runs either side of the mean.}
    \label{fig:dopinglevelscan}
\end{figure}
\fi

We can see that \CURTAINs not only outperforms \CATHODE~(local), but also approaches the performance of the Over-Idealised and supervised scenarios.
When using the full range outside of the signal regions to train \CATHODE~(full) the performance recovers and \CURTAINs is only able to match the performance at high levels of background rejection, as seen in \cref{fig:ROCandSIC_cathfull}.
However, this demonstrates that \CURTAINs is able to reach a higher level of performance when trained on lower numbers of events.

\iffiguresinbody
\begin{figure}[htpb]
    \centering 
    \includegraphics[width=\textwidth]{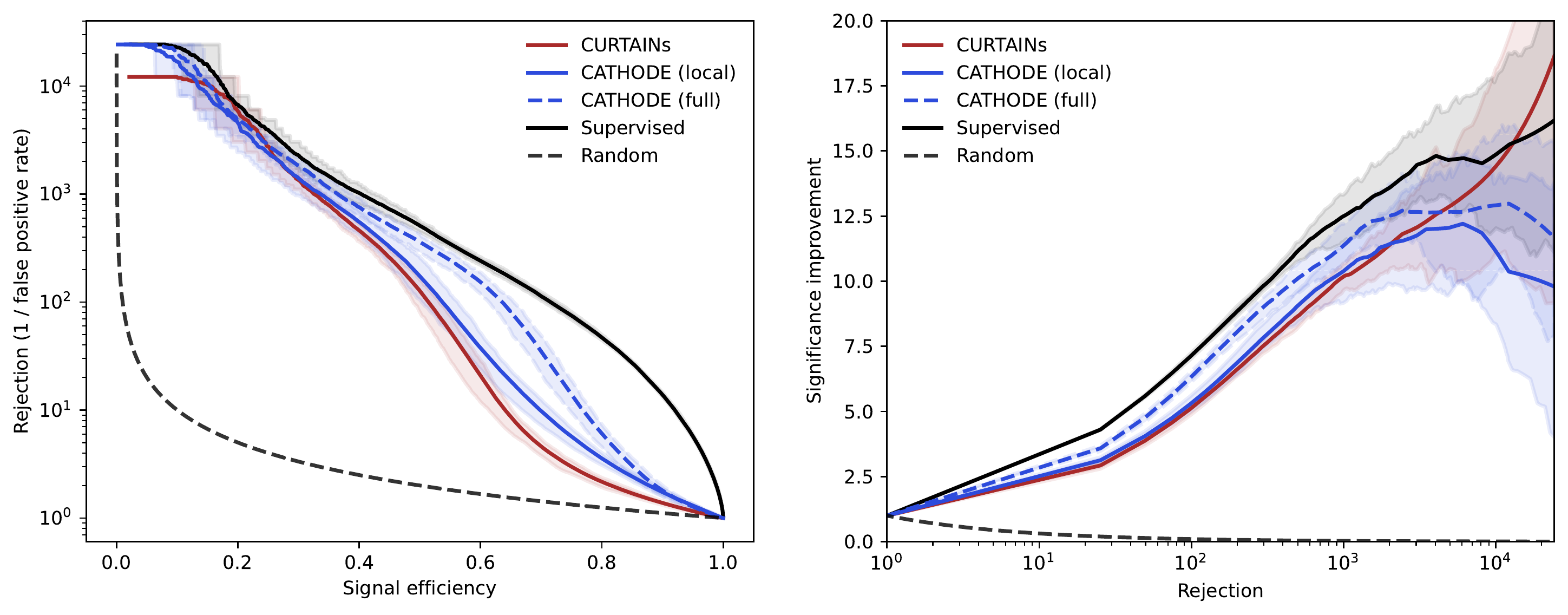}
    \caption{Background rejection as a function of signal efficiency (left) and signal improvement as a function of background rejection (right) for the \CURTAINs (red), \CATHODE~(local) (blue, solid), and \CATHODE~(full) (blue, dashed) background template models compared to a supervised classifier (black). The dashed \CATHODE~(full) model is trained using all data outside of the signal region, whereas the two solid lines are trained using the default 200~GeV side-bands.
    All classifiers are trained on the sample with 3000 injected signal events for the signal region 3300~$\leq m_{JJ} <$~3700~GeV.
    The lines show the mean value of fifty classifier trainings with different random seeds. The uncertainty encompasses 68\% of the runs either side of the mean. 
    }
    \label{fig:ROCandSIC_cathfull}
\end{figure}
\fi

\subsection{Application in a Sliding Window}
As it is not possible to know the location of the signal events when applying \CURTAINs to data, the real test of the performance and robustness of the method is in the sliding window setting.

Both \CURTAINs and \CATHODE~(local and full) are used to generate the background templates in a sliding window scan in the range 3000 to 4600 GeV, with steps of 200 GeV and equal 200~GeV wide signal regions. Classifiers are trained to separate the signal region data from the background template, and cuts are applied to retain 20\%, 5\% and 0.1\% of the background events.

These scans are performed for several levels of signal doping  and are shown in \cref{fig:bumphuntzero} for the case where there is no signal present, and in \cref{fig:bumphuntsignal} for doped samples with 500, 667, 1000 and 8000 injected signal events.
Each signal region is subdivided into two bins of equal width in $m_{JJ}$ for the plot.
The expected background is determined by multiplying the original yield of each bin by the chosen background retention factor.

In contrast to the \CWoLa bump hunt approach introduced in Ref.~\cite{cwolabump}, which uses the classifier trained in the signal region to apply a cut on all events on the invariant mass spectrum before performing a traditional bump hunt, we treat each signal region as an independent region and do not apply the classifiers outside of the regions in which they are trained.
This sliding window approach tests how the \CURTAINs and \CATHODE approaches perform as the sideband windows used to train the networks as well as the signal region transition between the presence of signal, to signal in one of the sidebands as well as the case where there is perfect alignment of signal in the signal region.
This approach does not test the ability of the trained classifiers to extrapolate outside of the values of invariant mass used to train them.
Were they to be applied outside of the respective regions it is expected sculpting of the invariant mass distribution would occur after applying cuts on the classifier due to the strong correlation between $\Delta R_{JJ}$ and $m_{JJ}$.

\iffiguresinbody
\begin{figure}[htpb]
    \centering
    \includegraphics[width=0.55\textwidth]{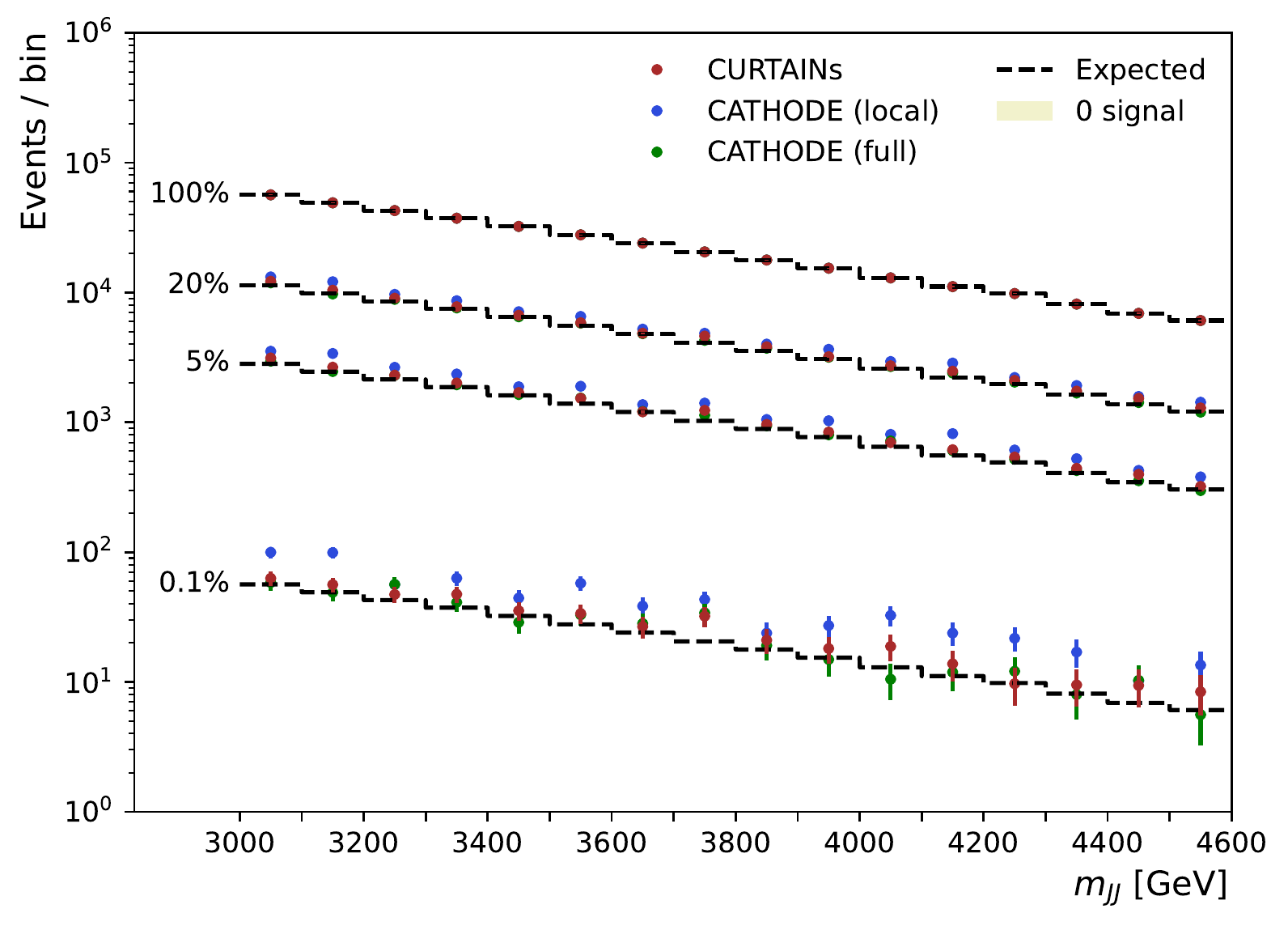}
    \caption{The dijet invariant mass for the range of signal regions probed in the sliding window, from 3300~GeV to 4600~GeV, for the case of zero doping.
    Each signal region is 200~GeV wide and split into two 100~GeV wide bins.
    The dashed line shows the expected background after applying a cut on classifier trained using the background predictions from the \CURTAINs (red), \CATHODE~(local) (blue) and \CATHODE~(full) (green) methods at specific background rejections. Three different cut levels are applied retaining 20\%, 5\% and 0.1\% of background events respectively. The cut values are calculated per signal region using the background template.}
    \label{fig:bumphuntzero}
\end{figure}
\fi

\iffiguresinbody
\begin{figure}[htpb]
    \centering
    \includegraphics[width=\textwidth]{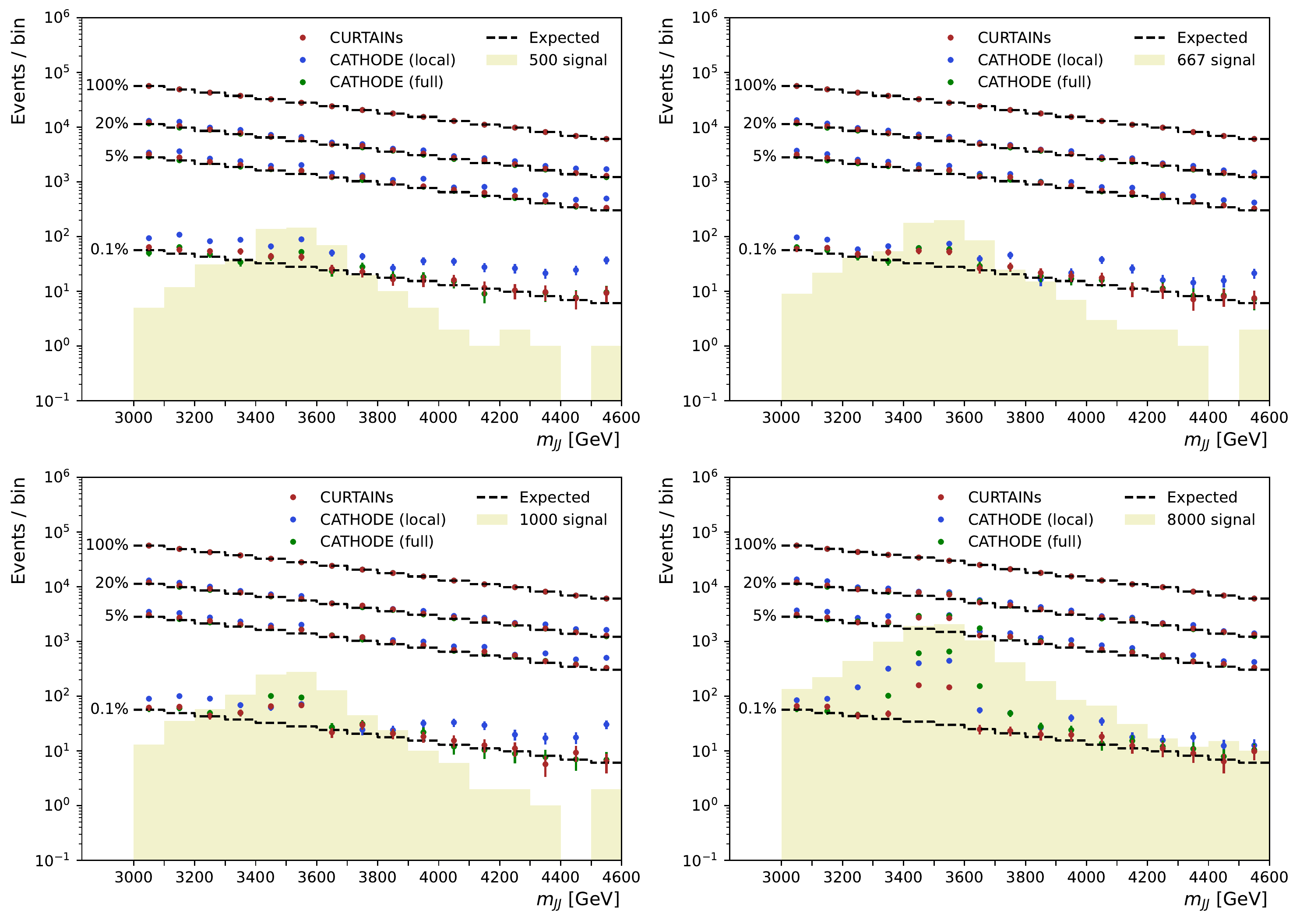}
    \caption{The dijet invariant mass for the range of signal regions probed in the sliding window, from 3300~GeV to 4600~GeV, for the case of samples doped with 500 (top left), 667 (top right), 1000 (bottom left) and 8000 (bottom right) signal events.
    Each signal region is 200~GeV wide and split into two 100~GeV wide bins .
    The dashed line shows the expected background after applying a cut on classifier trained using the background predictions from the \CURTAINs (red), \CATHODE~(local) (blue) and \CATHODE~(full) (green) methods at specific background rejections. Three different cut levels are applied retaining 20\%, 5\% and 0.1\% of background events respectively. The cut values are calculated per signal region using the background template.}
    \label{fig:bumphuntsignal}
\end{figure}
\fi

As can be seen from the sliding window scans in \cref{fig:bumphuntzero,fig:bumphuntsignal}, using \CURTAINs and \CATHODE~(full) we are able to correctly identify the location of the signal events for even reasonably low levels of signal. Where there are no or very few signal events the yields after each cut do not deviate too far from expected background.
The corresponding significance of excesses seen in each bin for three cuts on background efficiency are shown in \cref{fig:pvals}.
In the case where there is no signal injected into the sample, both \CURTAINs and \CATHODE~(full) have relatively low local excesses reaching a 4$\sigma$ deviation only in one bin at the 1\% background efficiency and not exceeding 3$\sigma$ deviations for tighter cuts when considering only statistical uncertainties on the yields in each bin.
\iffiguresinbody
\begin{figure}[htpb]
    \centering
    \includegraphics[width=\textwidth]{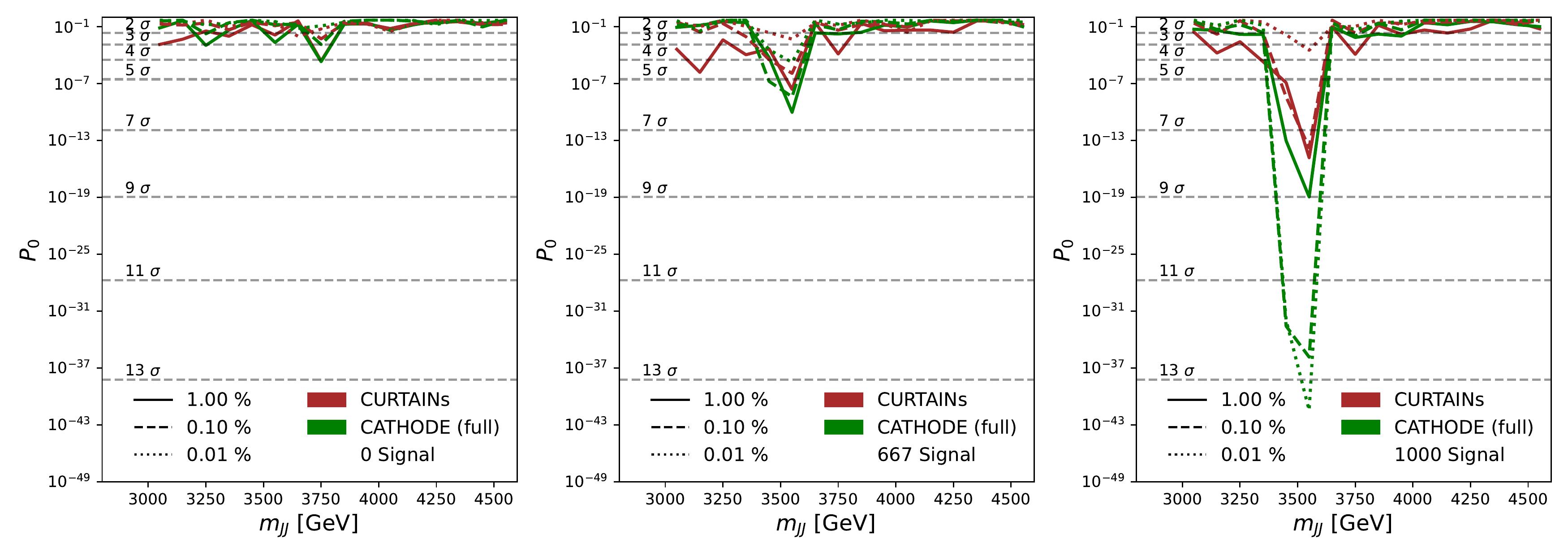}
    \caption{Measured excesses in each of the signal regions probed in the sliding window, from 3300~GeV to 4600~GeV, for the case of samples doped with 0 (left), 667 (middle), and 1000 (right) signal events.
    Each signal region is 200~GeV wide and split into two 100~GeV wide bins.
    The solid, dashed and dotted lines show the probability of the observed excesses $\left(p_{0}\right)$ over the background after applying a cut on classifier trained using the background predictions from the \CURTAINs (red) and \CATHODE~(full) (green) methods at 1\%, 0.1\%, and 0.01\% respectively. The cut values are calculated per signal region using the background template.}
    \label{fig:pvals}
\end{figure}
\fi

However, at lower levels of background rejection, significant local excesses over the expectation are observed. At 5\% background efficiency a maximum local deviation is observed at 4$\sigma$ for \CATHODE~(full) and 5$\sigma$ for \CURTAINs. 
In the presence of signal both \CURTAINs and \CATHODE~(full) have observed excesses at the signal mass peak for each cut level for the lower levels of signal events, with \CATHODE~(full) approaches results in a more prominent excess.

The \CATHODE~(local) approach yields an excess across the whole spectrum in the absence of signal and for all levels of injected signal. However, it also finds an excess under the signal peak in the cases where signal is injected, which at higher levels of signal exceeds that found by \CURTAINs.

In an analysis a systematic excess over the expectation calculated from the original yields per bin would not necessarily be problematic, as the expectation could be determined from a side band fit in $m_{JJ}$ after applying the cut. 
Additionally, these values do not take any systematic uncertainties into account and only consider statistical uncertainties on the number of events passing each cut from the yields.


Although the ability to isolate the signal events when using \CURTAINs in the window scan decreases at low numbers of signal events and signal purity, this is also seen for both idealised cases in \cref{fig:dopinglevelscan} and suggests this as rather an area where the classifier architecture and anomaly detection method need to be optimised.
The performance of \CURTAINs in this setting could also be further improved by optimising the binning used in the sliding window, and the number of subdivisions within each signal region.

\section{Conclusions}
In this paper we have proposed a new method, \CURTAINs, for use in weakly supervised anomaly detection which can be used to extend the sensitivity of bump hunt searches for new resonances. This method stays true to the bump hunt approach by remaining completely data driven, and with all templates and signal extraction performed on a local region in a sliding window configuration. 

\CURTAINs is able to produce a background template in the signal region which closely matches the true background distributions.
When applied in conjunction with anomaly detection techniques to identify signal events, \CURTAINs matches the performance of an idealised setting in which the background template is defined using background events from the signal region and approaches the performance of an oversampled idealised setting. It also does not produce spurious excesses in the absence of signal events.

As real data points are used with the \CURTAINs transformer to produce the background template, we avoid problems which can arise from sampling a prior distribution leading to non perfect agreement over distributions of features and their correlations. By conditioning the transformation on the difference in input and target $m_{JJ}$, we also avoid the need to interpolate or extrapolate outside of the values seen in training.
Using this approach we see \CURTAINs is able to reach similar levels of performance in comparison to state-of-the-art methods.
\CURTAINs delivers this performance even when using much less training data, as seen when using side-bands as opposed to the full data distribution outside of the signal region. 

Another key advantage of \CURTAINs over other proposed techniques is the ability to apply it to validation regions.
By transforming the side-bands data to other regions than the signal region, validation regions can be defined in which the transformer and classifier architectures can be optimised on real data. Here the \CURTAINs transformer can be validated and optimised by ensuring the agreement between the transformed data and target data distributions is as close as possible, and the classifier architecture can be optimised to make sure it does not pick up on residual differences between transformed and target data. In this paper only the former optimisation procedure was performed, with the classifier architecture instead chosen for its robustness to variability in initial conditions.

However, care must be taken to optimise the width of the signal region when training the \CURTAINs model to make sure that the signal to background ratio is not constant across the side-band and signal regions.

It may be possible to extend \CURTAINs to extrapolation tasks, where a model would be trained on one control region and applied to all other regions. This could allow one model to be trained per bump hunt, or a model could be trained to extrapolate to the tails of distributions, allowing these regions to be probed in a model independent fashion. Thanks to its performance and ability to be applied to a sliding window fit, \CURTAINs is simple to apply to current sliding window fits and should bring significant gains in sensitivity 
in the search for new physics at the LHC and other domains.

    \section*{Acknowledgements}
    The authors would like to acknowledge funding through the SNSF Sinergia grant called Robust Deep Density Models for High-Energy Particle Physics and Solar Flare Analysis (RODEM) with funding number CRSII$5\_193716$, and the SNSF project grant 200020\_181984 called Exploiting LHC data with machine learning and preparations for HL-LHC.

    The authors would also like to thank Matthias Schlaffer, our resident \CATHODE Guru, for his invaluable input in establishing a reliable baseline for comparisons and useful discussions, and Knut Zoch for input on the initial studies and samples used. Both Knut and Matthias are also thanked for their feedback on this manuscript.
Further thanks go to Tobias Quadfasel, Manuel Sommerhalder and David Shih for fruitful discussions and helping improve harmonisation with the implementation of the \CATHODE model.

    \phantomsection
    \addcontentsline{toc}{chapter}{References}
    \printbibliography[title=References]

    \clearpage
    \appendix
    
    \section{Additional Tables}

    \begin{table}[htpb]
      \caption {The performance of the CURTAINs transformer when trained on different amounts of training data. For all models the sidebands, signal region and outer-bands are 200 GeV wide, with the model trained on \SBone (3200~$\leq m_{JJ} <$~3400~GeV) and \SBtwo (3600$~\leq m_{JJ} <$~3800~GeV). Validation of performance is measured using a classifier trained on the transformed data against the real data in the target region and quantified by the area under the ROC curve.
      For 100\% of the available statistics, there are around 80,000 training examples in \SBone and 45,000 training examples in \SBtwo.
      In all cases there are no signal events injected into the training sample.
      The model architecture optimised for the full available training dataset has been used without any optimisation for all levels.
      The training time scales linearly with the size of the training dataset, with the full available statistics requiring 6 hours to train for 1000 epochs on an Nvidia RTX 3080 GPU.
      The performance decreases with the available training data, with the outer-bands most impacted. The performance stays around the level of 0.50--0.52 for all transformations down to 25\% of the available training statistics, though a noticeable drop in performance is observed between the two side-bands  at 50\%.
      }
      \label{tab:curtains_stats}
      \centering
      \begin{tabular}{r c c c c | c}
          \hline
          \hline
          Training statistics & \SBone\!\!$\rightarrow$\SBtwo & \SBtwo\!$\rightarrow$\SBone & \SBone\!\!$\rightarrow$\OBone & \SBtwo\!$\rightarrow$\OBtwo & \thead{\SBone\!\!$\rightarrow$SR \\ $\cup$\\ \SBtwo\!$\rightarrow$ SR}\\
          \hline
          
          100\%  & 0.504 & 0.504 & 0.519 & 0.512 & 0.509   \\
          75\%   & 0.508 & 0.509 & 0.518 & 0.509 & 0.505   \\
          50\%   & 0.518 & 0.519 & 0.518 & 0.521 & 0.506   \\
          25\%   & 0.518 & 0.520 & 0.512 & 0.516 & 0.505   \\
          10\%   & 0.533 & 0.534 & 0.527 & 0.523 & 0.508   \\
          5\%    & 0.530 & 0.532 & 0.534 & 0.527 & 0.511   \\
          \hline\hline
      \end{tabular}
  \end{table}

\clearpage
\section{Additional Figures}



    \begin{figure}[htpb]
        \centering
        \includegraphics[width=0.49\textwidth]{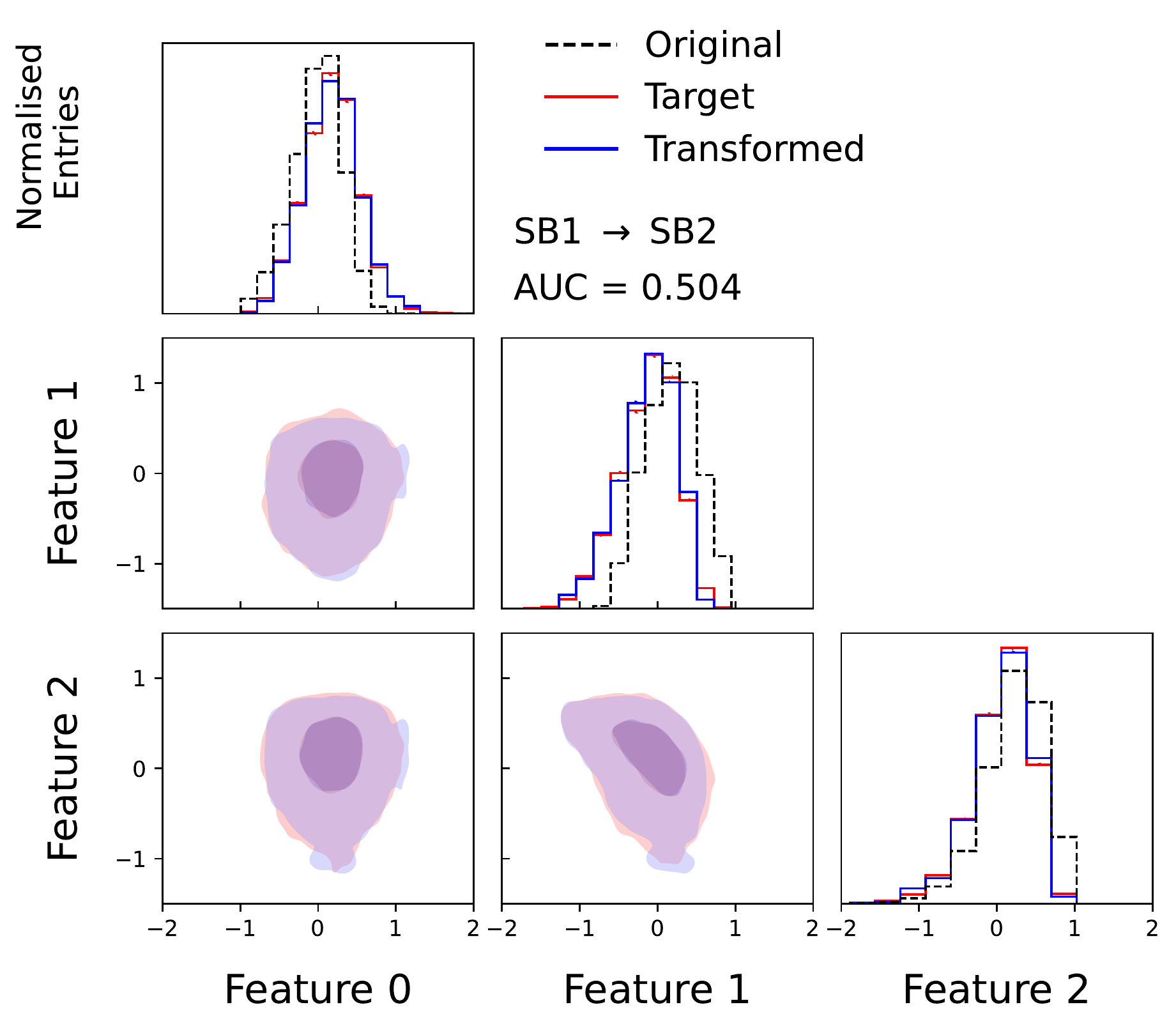}
        \includegraphics[width=0.49\textwidth]{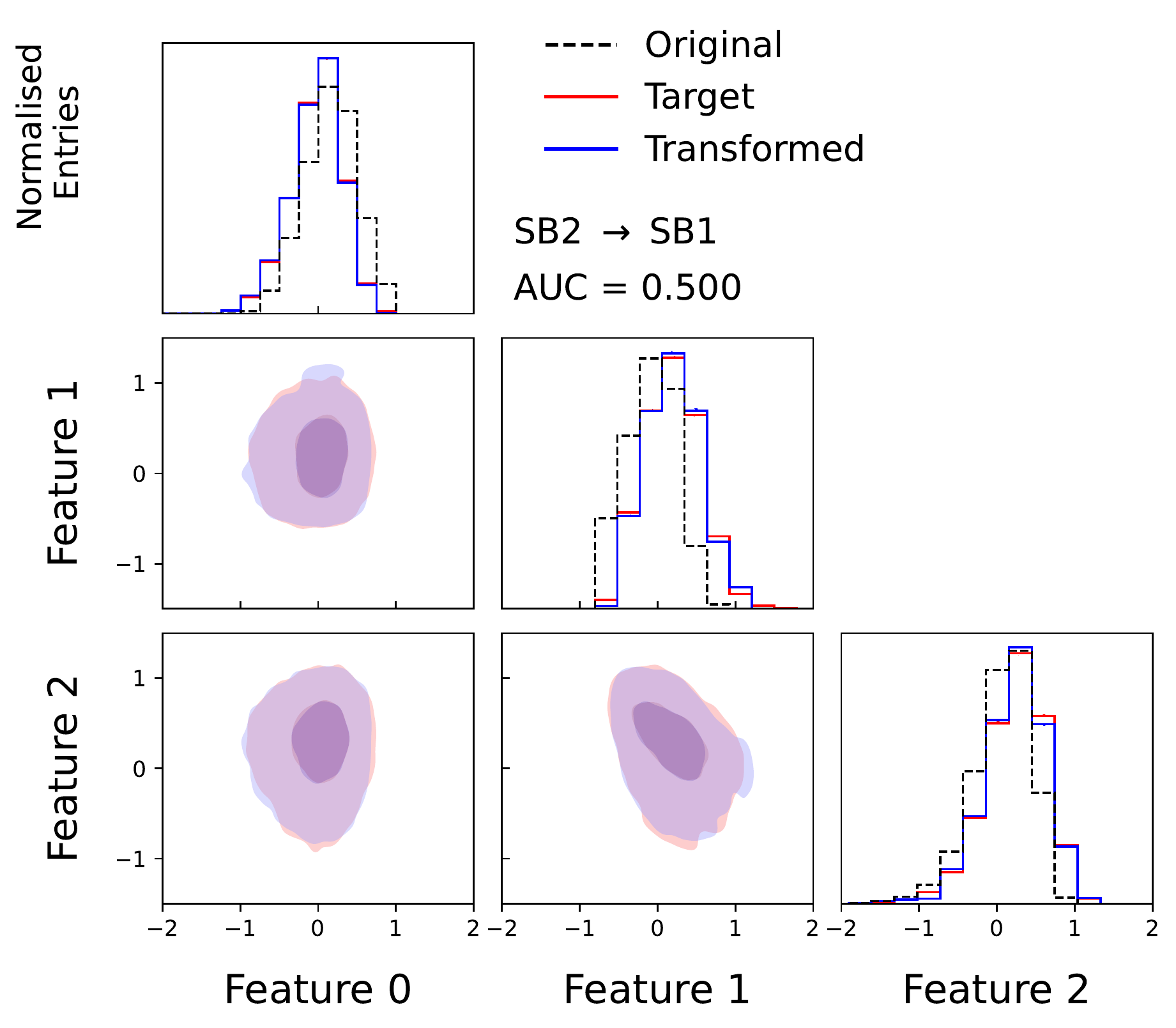}
        \caption{Input, target and transformed data distributions for a toy example with enhanced correlation between the input variables and the invariant mass. Features 0 and 1 are linearly correlated to $m_{JJ}$ whilst and Feature 3 is proportional to the inverse cube of $m_{JJ}$.
        In comparison to the architecture optimised for the nominal features, the size of each coupling layer has been increased to four residual blocks, each with two hidden layers of 64 nodes.
        Distributions are shown for transforming data from \SBone to \SBtwo (left) and \SBtwo to \SBone (right), with the model trained on \SBone (3200~$\leq m_{JJ} <$~3400~GeV) and \SBtwo (3600$~\leq m_{JJ} <$~3800~GeV). The data from \SBone (\SBtwo) is transformed with a forward (inverse) pass of the \CURTAINs model into the target region.
        The diagonal elements show the individual features with the off diagonal elements showing a contour plot between the two observables for the transformed and trained data.}
        \label{fig:curtains_nosignal_SBs}
    \end{figure}

    \begin{figure}[htpb]
        \centering
        \includegraphics[width=0.49\textwidth]{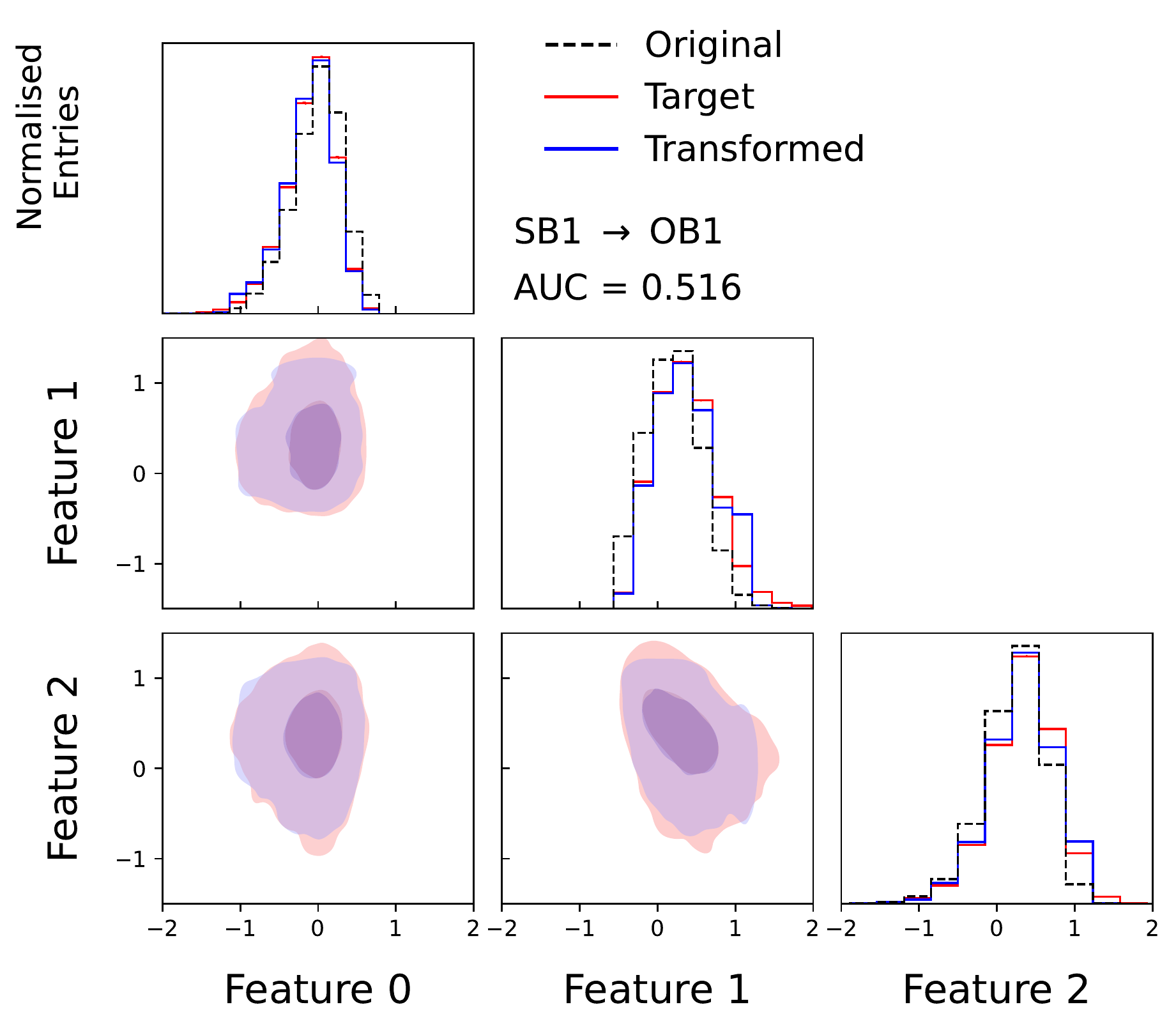}
        \includegraphics[width=0.49\textwidth]{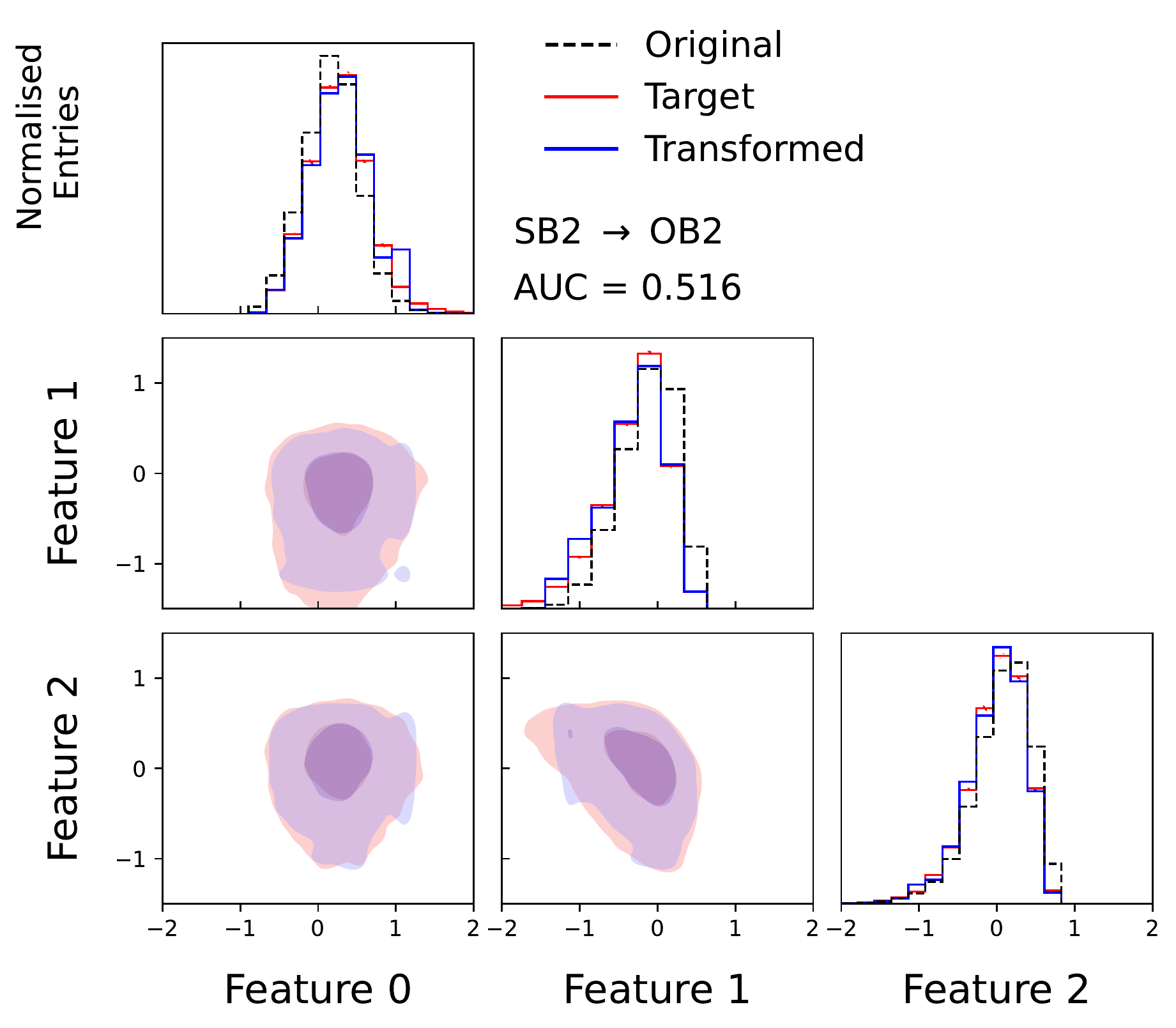}
        \caption{Input, target and transformed data distributions for a toy example with enhanced correlation between the input variables and the invariant mass. Features 0 and 1 are linearly correlated to $m_{JJ}$ whilst and Feature 3 is proportional to the inverse cube of $m_{JJ}$. 
        In comparison to the architecture optimised for the nominal features, the size of each coupling layer has been increased to four residual blocks, each with two hidden layers of 64 nodes.
        Distributions are shown for transforming data from \SBone to \OBone (left) and \SBtwo to \OBtwo (right), with the model trained on \SBone (3200~$\leq m_{JJ} <$~3400~GeV) and \SBtwo (3600$~\leq m_{JJ} <$~3800~GeV), with \OBone and \OBtwo defined as 200~GeV wide windows directly next to \SBone and \SBtwo away from the signal region.
        The data from \SBone (\SBtwo) is transformed with an inverse (forward) pass of the \CURTAINs model into the target region.
        The diagonal elements show the individual features with the off diagonal elements showing a contour plot between the two observables for the transformed and trained data.}
        \label{fig:curtains_nosignal_OBs}
    \end{figure}

    \begin{figure}[htpb]
        \centering
        \includegraphics[width=0.66\textwidth]{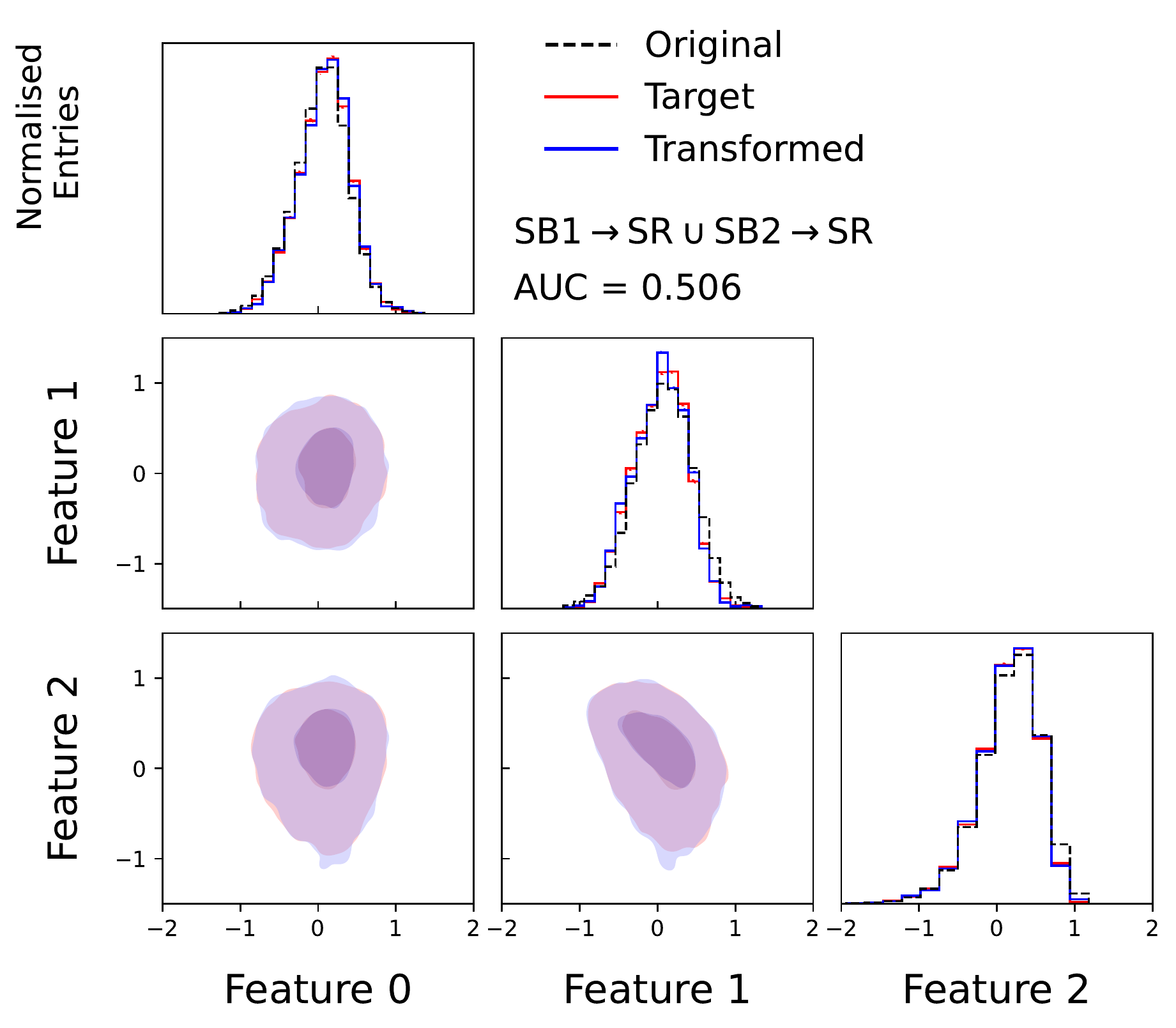}
        \caption{Input, target and transformed data distributions for a toy example with enhanced correlation between the input variables and the invariant mass. Features 0 and 1 are linearly correlated to $m_{JJ}$ whilst and Feature 3 is proportional to the inverse cube of $m_{JJ}$. 
        In comparison to the architecture optimised for the nominal features, the size of each coupling layer has been increased to four residual blocks, each with two hidden layers of 64 nodes.
        Distributions are shown for transforming data from \SBone and \SBtwo to the signal region to create the background template, with the model trained on \SBone (3200~$\leq m_{JJ} <$~3400~GeV) and \SBtwo (3600$~\leq m_{JJ} <$~3800~GeV). The data from \SBone (\SBtwo) is transformed with a forward (inverse) pass of the \CURTAINs model into the target region.
        The diagonal elements show the individual features with the off diagonal elements showing a contour plot between the two observables for the transformed and trained data.}
    \end{figure}

    \begin{figure}[htpb]
      \centering
      \includegraphics[width=0.49\textwidth]{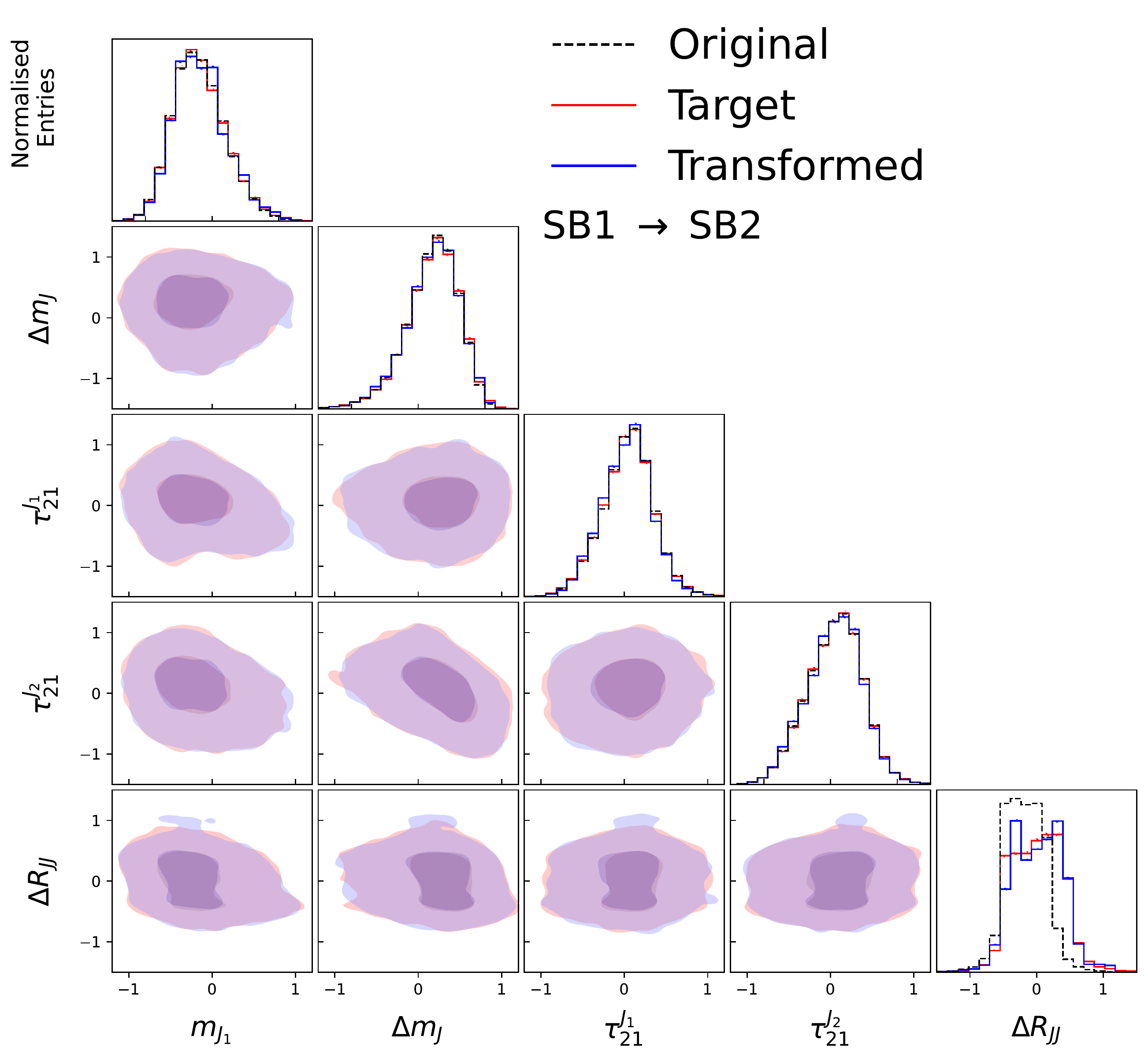}
      \includegraphics[width=0.49\textwidth]{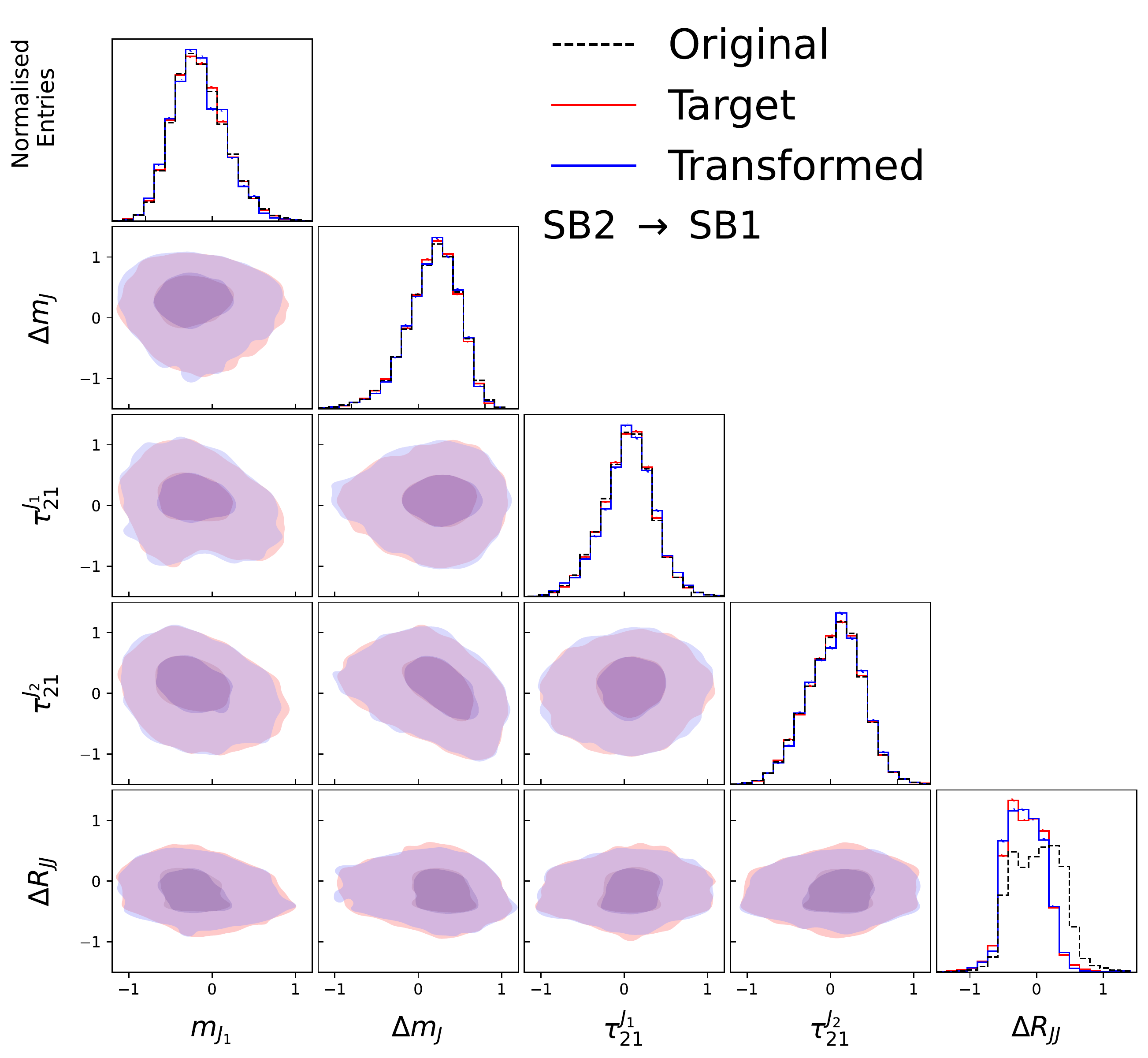}
      \caption{Input, target and transformed data distributions for the base variable set with the addition of $\Delta R_{JJ}$, for transforming data from \SBone to \SBtwo (left) and \SBtwo to \SBone (right), with the model trained on \SBone (3200~$\leq m_{JJ} <$~3400~GeV) and \SBtwo (3600$~\leq m_{JJ} <$~3800~GeV). The data from \SBone (\SBtwo) is transformed with a forward (inverse) pass of the \CURTAINs model into the target region.
      The diagonal elements show the individual features with the off diagonal elements showing a contour plot between the two observables for the transformed and trained data.
      Only 25\% of the available data has been used to train the \CURTAINs model.
      The model architecture optimised for the full available training dataset has been used without any optimisation.}
  \end{figure}

  \begin{figure}[htpb]
      \centering
      \includegraphics[width=0.49\textwidth]{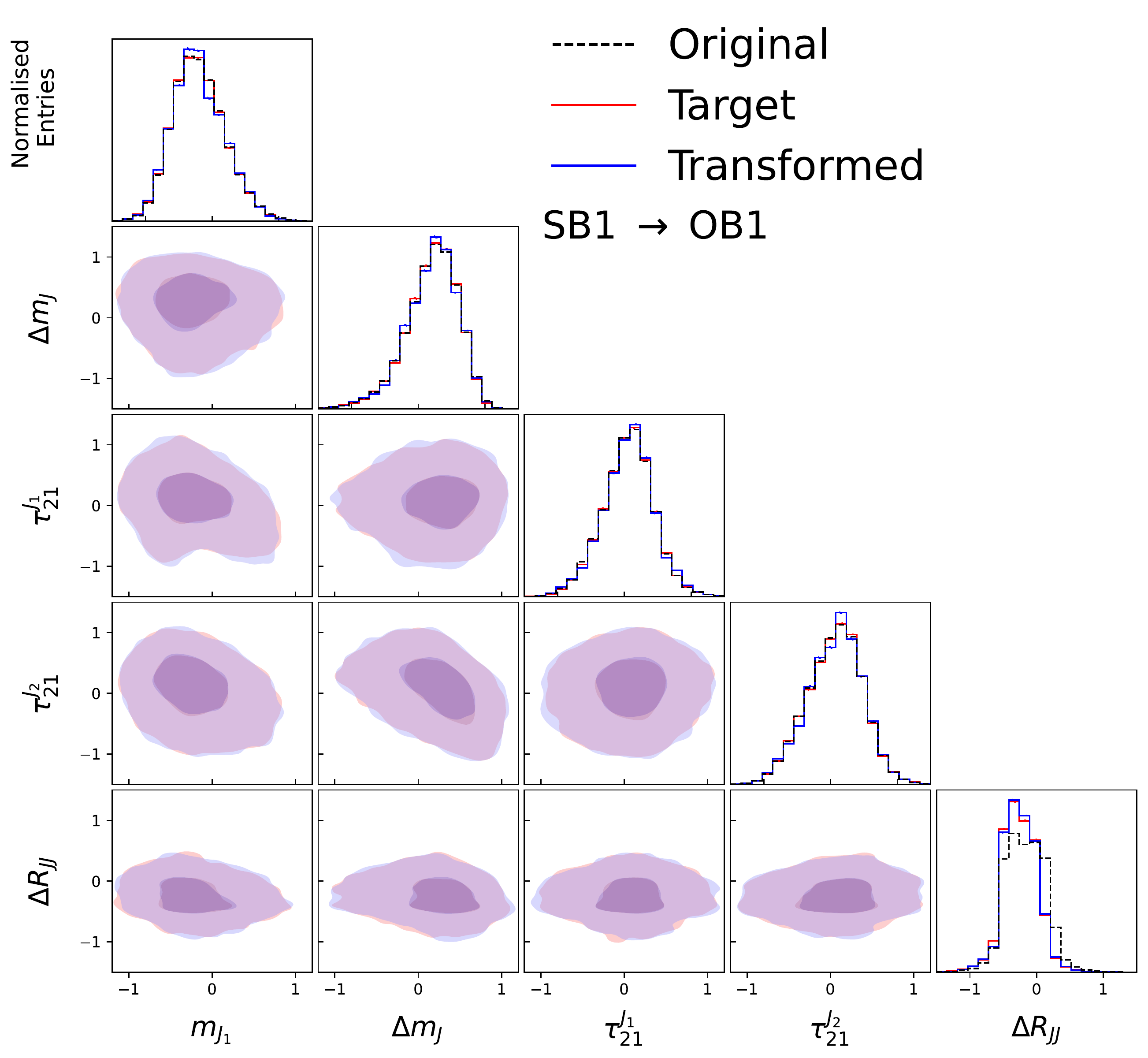}
      \includegraphics[width=0.49\textwidth]{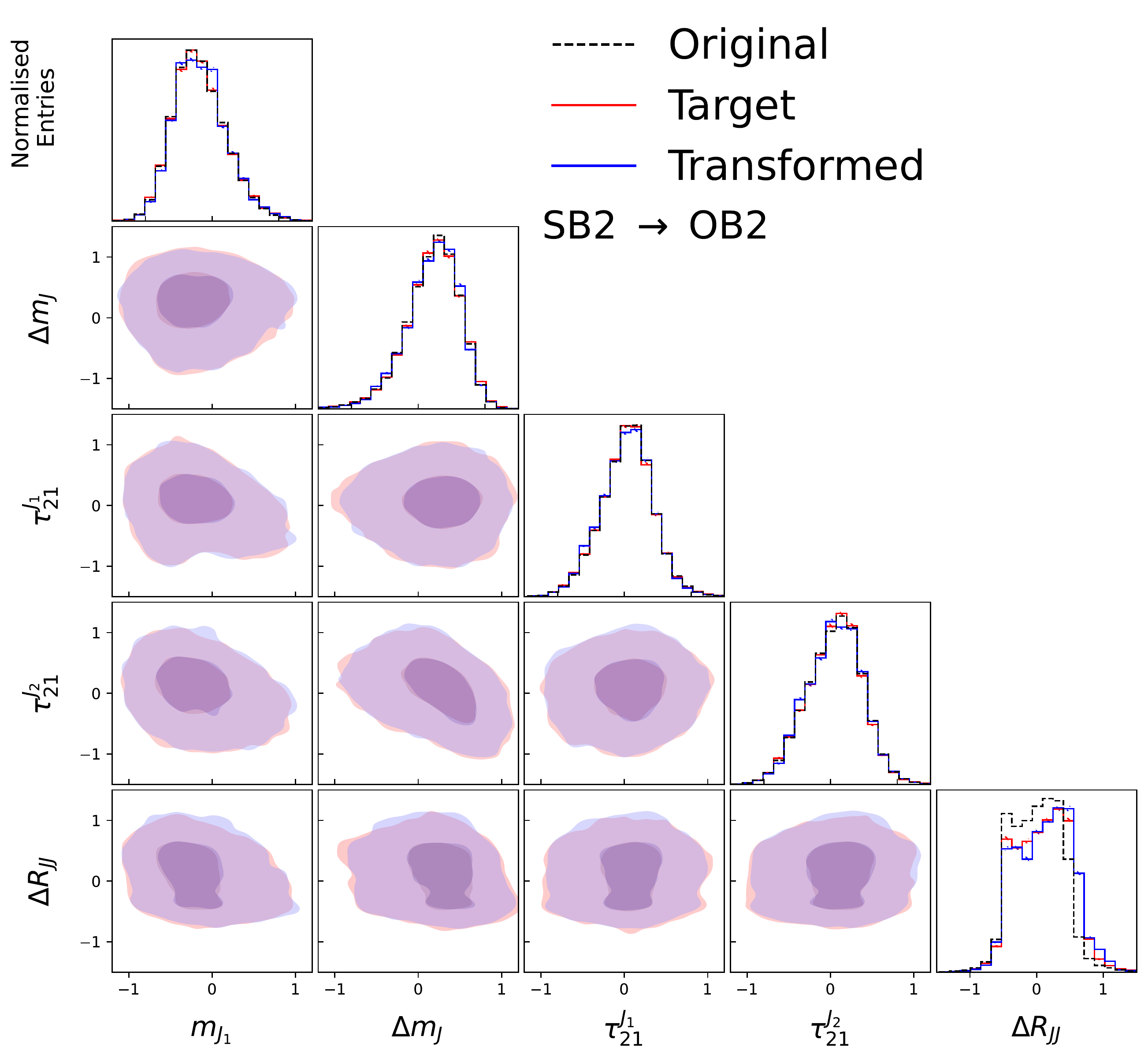}
      \caption{Input, target and transformed data distributions for the base variable set with the addition of $\Delta R_{JJ}$, for transforming data from \SBone to \OBone (left) and \SBtwo to \OBtwo (right), with the model trained on \SBone (3200~$\leq m_{JJ} <$~3400~GeV) and \SBtwo (3600$~\leq m_{JJ} <$~3800~GeV), with \OBone and \OBtwo defined as 200~GeV wide windows directly next to \SBone and \SBtwo away from the signal region.
      The data from \SBone (\SBtwo) is transformed with an inverse (forward) pass of the \CURTAINs model into the target region.
      The diagonal elements show the individual features with the off diagonal elements showing a contour plot between the two observables for the transformed and trained data.
      Only 25\% of the available data has been used to train the \CURTAINs model.
      The model architecture optimised for the full available training dataset has been used without any optimisation.}
  \end{figure}

  \begin{figure}[htpb]
      \centering
      \includegraphics[width=0.66\textwidth]{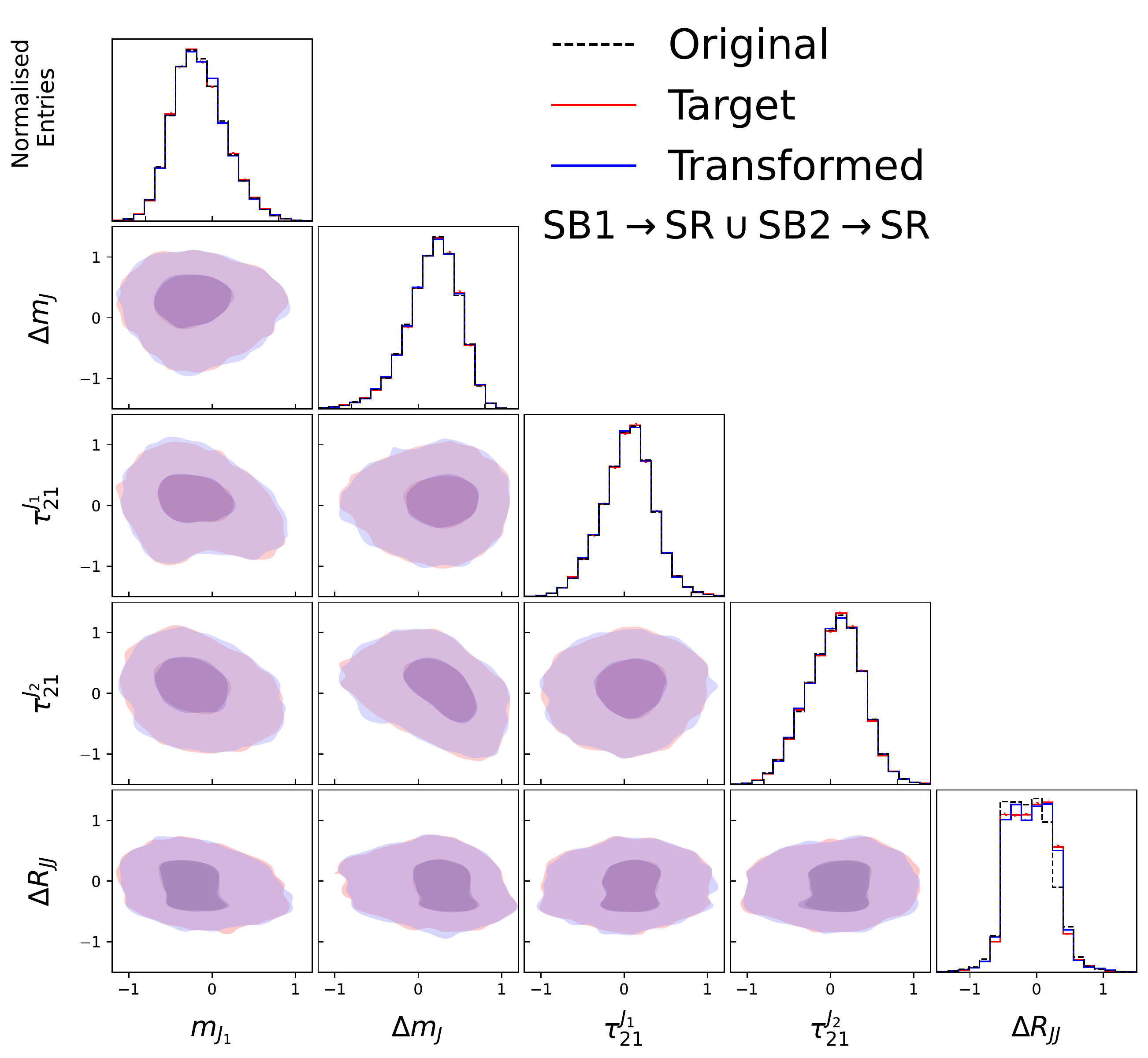}
      \caption{Input, target and transformed data distributions for the base variable set with the addition of $\Delta R_{JJ}$, for transforming data from \SBone and \SBtwo to the signal region to create the background template, with the model trained on \SBone (3200~$\leq m_{JJ} <$~3400~GeV) and \SBtwo (3600$~\leq m_{JJ} <$~3800~GeV). The data from \SBone (\SBtwo) is transformed with a forward (inverse) pass of the \CURTAINs model into the target region.
      The diagonal elements show the individual features with the off diagonal elements showing a contour plot between the two observables for the transformed and trained data.
      Only 25\% of the available data has been used to train the \CURTAINs model.
      The model architecture optimised for the full available training dataset has been used without any optimisation.}
  \end{figure}

  \begin{figure}[htpb]
      \centering
      \includegraphics[width=\textwidth]{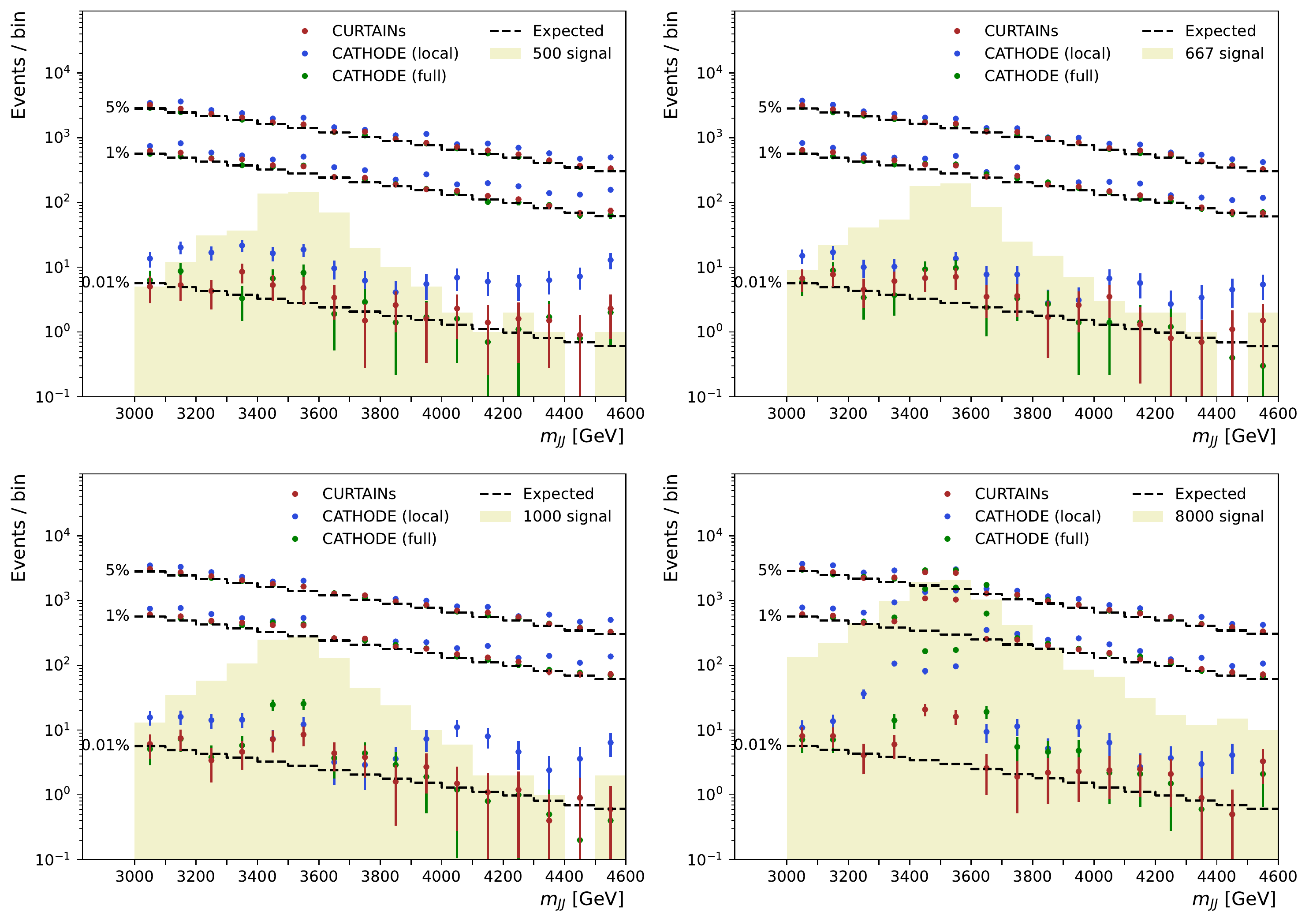}
      \caption{The dijet invariant mass for the range of signal regions probed in the sliding window, from 3300~GeV to 4600~GeV, for the case of samples doped with 500 (top left), 667 (top right), 1000 (bottom left) and 8000 (bottom right) signal events.
      Each signal region is 200~GeV wide and split into two 100~GeV wide bins .
      The dashed line shows the expected background after applying a cut on classifier trained using the background predictions from the \CURTAINs (red), \CATHODE~(local) (blue) and \CATHODE~(full) (green) methods at specific background rejections. Three different cut levels are applied retaining 5\%, 1\% and 0.01\% of background events respectively. The cut values are calculated per signal region using the background template.}
      \label{fig:bumphuntsignal_zoom}
  \end{figure}

\end{document}